\documentstyle[preprint,tighten,epsfig,aps,floats]{revtex}

\newcommand{\fd}{fluc\-tu\-a\-tion-dis\-si\-pa\-tion }
\newcommand{\be}{\begin{equation}}
\newcommand{\bea}{\begin{eqnarray}}
\newcommand{\ee}{\end{equation}}
\newcommand{\eea}{\end{eqnarray}}
\newcommand{\eq}{{\rm eq}}
\newcommand{\dpar}{\partial}
\renewcommand{\o}{{\mathcal O}}
\newcommand{\ot}{{\widetilde{\mathcal O}}}
\newcommand{\pt}{{\tilde\varphi}}
\newcommand{\p}{{\varphi}}
\newcommand{\scal}{{\rm scal}}
\newcommand{\can}{{\rm can}}
\newcommand{\li}{{\rm Li}_2}
\def\en{\varepsilon}
\def\e{\epsilon}
\def\H{{\cal H}} 
\begin{document}

\draft

\title{On the definition of a unique effective temperature for non-equilibrium critical systems}
\author{Pasquale Calabrese${}^{1}$ and Andrea Gambassi${}^{2,3}$}
\address{$^1$Rudolf Peierls Centre for Theoretical Physics, 1 Keble Road,
Oxford OX1 3NP, United Kingdom.}
\address{$^{2}$ Max-Planck-Institut f\"ur Metallforschung, Heisenbergstr. 3,
D-70569 Stuttgart, Germany.}
\address{$^{3}$ Institut f\"ur Theoretische und Angewandte Physik,
Universit\"at Stuttgart, Pfaffenwaldring 57, D-70569 Stuttgart, Germany.}

\date{\today}

\maketitle

\begin{abstract}

We consider the problem of the 
definition of an effective temperature via the long-time limit of the
\fd ratio $X^\infty$ after a quench from the disordered state 
to the critical point of an $O(N)$ model with dissipative dynamics.
The scaling forms of the response and correlation functions
of a generic observable $\o(t)$ are derived from the solutions of the 
corresponding Renormalization Group equations.
We show that within the Gaussian approximation all the local 
observables have the same $X^\infty_\o$, allowing for a definition of a unique 
effective temperature. This is no longer the case when fluctuations
are taken into account beyond that approximation, as shown
by a computation up to the first order in the $\e$-expansion for two 
quadratic observables.
This implies that, contrarily to what often conjectured,  
a unique effective temperature can not be defined for this 
class of models.

\end{abstract}

\pacs{PACS Numbers: 05.10.Cc, 05.10.Gg, 64.60.Ht, 75.40.Gb}

% ========================= BODY =========================
%\narrowtext

\section{Introduction}
\label{intr}

The non-equilibrium dynamics of physical systems is 
one of the most challenging problems in statistical
mechanics. Equilibrium statistical
mechanics has been probably one of the most important achievements
during the last century. 
On the other hand, in nature equilibrium is more an
exception rather than a rule. 
In view of that many efforts are currently aiming at achieving a 
coherent theoretical  picture of non-equilibrium phenomena: Indeed
many real systems persist out-of-equilibrium practically forever.
One example is naturally provided by 
slow-relaxing systems, such as structural glasses and 
spin-glasses, whose equilibration times sometimes exceed any reasonable
experimental time scale (in fact they can even evolve on the scale of
geological eras). 
The high degree of complexity of such systems 
makes the description of their dynamics 
really awkward, since apparently all the history of 
a sample has to be known in order to predict its evolution. Conversely, a
useful theory should be able to provide a description of the behavior
of the system in terms of few and essential effective parameters.
The effective temperature, defined on the basis of \fd relations
between correlation and response functions \cite{ckp-97}, has been proved 
very fruitful in this direction (at least for mean-field models) and 
it is currently under intensive experimental investigation \cite{exp}.

To introduce the concept let us consider the following experiment.
Prepare a system (e.g., a glass, a ferromagnet, etc.) in 
an equilibrium state corresponding to a 
high temperature $T_0$ (where ``high'' means, here, 
greater than any critical or glass transition temperature). 
At time $t=0$, quench the system to some 
temperature $T<T_0$ by taking it into contact with a thermal
bath at temperature $T$, and let it evolve. 
On a general basis, one expects that the relaxation towards the equilibrium 
state corresponding to $T$ is characterized by two different regimes: 
(A) a transient one with non-equilibrium evolution, for $t<t_{\rm \eq}(T)$, 
and (B) a stationary regime with equilibrium evolution for $t>t_{\rm \eq}(T)$,
where $t_{\rm eq}(T)$ is some characteristic equilibration time of the system.
During (A) the behavior of the system is expected to depend
on the specific initial conditions 
and both time-reversal symmetry and time-translation invariance 
are broken, while they are recovered in regime (B): The dynamics
of fluctuations is given by the ``equilibrium'' one.
We are concerned here with those 
systems for which the regime (B) is never achieved during  
experimental times, i.e., for all practical purposes, $t_{\rm \eq} = \infty$.
In this case, the standard concepts of equilibrium statistical mechanics do 
not apply and in particular two-time quantities, 
such as the response  
$R_{\o }(t,s)= \left. \langle \frac{\delta\o(t)}{\delta h_\o(s)}
\rangle\right|_{h_\o=0}$
and the correlation functions
$C_\o(t,s) = \langle \o(t)\o(s)\rangle$ ($\o$ is some observable and
$h_\o$ its conjugated field, we will be more explicit later on)
depend separately on $s$~(usually called the
``age'' of the system, being the time spent in the phase with 
$t_{\rm eq}=\infty$) and $t$, even for long times.
This behavior is usually referred to as aging \cite{struik,review}. 
Such a useful quantity as the temperature $T$ 
of the system is not defined in this genuine non-equilibrium
regime. On the other hand one can address the question whether some
effective temperature $T_{\rm eff}$ (in general different from $T$) 
can be still defined and used to understand the physics of the system.

In equilibrium [regime (B)], 
correlation and response functions depend only on time 
differences and the \fd theorem (FDT) states that
\be
T R_\o(t-s)= \partial_s C_\o(t-s)\,,
\label{FDT}
\ee
[here and in the
following we assume $t>s$ given that causality implies $R_\o(t<s,s)=0$]
where $T$ is expressed in units $k_B=1$. Whatever the regime is, one
can always define the so-called \fd ratio (FDR) as \cite{ck-93}
\be
X_\o(t,s)=\frac{T\, R_{\o}(t,s)}{\dpar_s C_{\o}(t,s)} \;.
\label{FDRgen}
\ee
As a consequence of Eq.~(\ref{FDT}), $X_\o(t,s)=1$ whenever
$t>s\gg t_{\rm eq}(T)$.
Some sort of effective temperature \cite{ckp-97} in the 
aging regime [$t_{\rm eq}(T) = \infty$] 
can be defined via the long-time limit of $X_\o(t,s)$
\be
X^\infty_\o= \lim_{s\rightarrow\infty}
\lim_{t\rightarrow\infty}X_\o(t,s)\;,
\label{FDR}
\ee
through the relation $T^\o_{\rm eff}=T/X_\o^\infty$ (that reduces to
the thermodynamic temperature of the thermal bath when equilibrium is asymptotically reached).
Obviously this definition has to be regarded as formal as long as one
is not able to establish a link between $T_{\rm eff}$ and some
thermodynamic properties. Nevertheless, in Ref. \cite{ckp-97} it has been 
shown that $T_{\rm eff}$ plays the same role as the 
thermodynamic temperature, in the sense that 
it determines the direction of heat flows and acts as a criterion for 
thermalization. Moreover, a thermometer coupled to the observable $\o$ 
measures (on a proper time scale) 
the temperature $T_{\rm eff}^\o$ \cite{ckp-97}. 
We also mention that it has been argued that 
$X_{\o}(t,s)$ establishes a bridge between the dynamically inaccessible 
equilibrium state and the asymptotic dynamics for large times \cite{fmpp-98}.
(See also Refs. \cite{cugl-02,cr-03} for a discussion of 
other properties of $X_{\o}(t,s)$ and $T^\o_{\rm eff}$.)

It has been stressed several times in the literature 
(see, e.g., Ref.~\cite{sfm-02}) that the effective temperature can 
be of interest in order to devise some thermodynamics for the system
provided that 
its value is independent of the observable used to define it.
This has been explicitly verified for infinite-range  
(mean-field) glass models \cite{ckp-97}.
Beyond these cases, the observable dependence of $T_{\rm eff}^\o$
has been investigated analytically for the trap model \cite{fs-02},
for the one-dimensional Ising model \cite{mbgs-03,ms-04}, and for the 
$d$-dimensional spherical model \cite{s04}, whereas numerical studies
have addressed the problem for supercooled liquids \cite{bb-02} and for the 
two-dimensional Ising model \cite{mbgs-03}. 

In this paper we discuss the problem of the observable independence of the 
FDR (and consequently of $T^\o_{\rm eff}$) in a simpler 
(compared to glasses) class of slow-relaxing systems: 
Critical systems quenched from a high-temperature phase to the critical point
and evolving according to a purely dissipative dynamics. 
%%%%%%%%
%%%%%%%%
In fact, soon after the introduction of the FDR Eq. (\ref{FDRgen}), 
it was pointed out \cite{ckp-94} that also these systems display
slow-relaxation (due to $t_{\rm eq} \sim \xi^z$, 
where $\xi$ is the correlation length, diverging at 
the critical point, and $z$ the dynamical critical exponent
\cite{HH})
and aging. The FDR of the order parameter was then determined for a Gaussian
model and for the random walk.
Subsequent  analytical and numerical calculations on realistic 
models confirmed and generalized this picture \cite{lz-00,gl-00i,fins-01,gl-00c,bhs-01,hpgl-01,henkel-02,gl-02,cg-02a1,cg-02a2,cg-02rim,cg-03,c-03,sdc-03,ak-04,c-04,ak-04b,sl-03,ph-02,hu-03,hs-04,ph-04,gkr-04,p-04,ms-04b,cont,mbgs-03,ms-04,s04}
(for a review, see Ref. \cite{cg-rev}).
For our analysis we can take advantage of
the powerful tools of Renormalization Group (RG) and field theory to
provide analytical predictions for some FDR's in an $\e$-expansion, 
where $\e=4-d$ and $d$ is the spatial dimensionality of the system. 
This kind of study allows us to clarify once and
for all whether a {\it unique} effective temperature can be defined for such 
systems in the long-time limit.

The outline of the paper is the following.
In Sec. \ref{sec2} we introduce the model and the field-theoretical approach
to non-equilibrium dynamics. We derive, by means of RG equations, the 
scaling forms of two quadratic (in the order parameter) observables. 
In Sec. \ref{sec3} we argue that the FDR $X_\o^\infty$ 
(and thus $T^\o_{\rm eff}$) does
not depend on the specific observable $\o$ within the
Gaussian (mean-field) approximation. 
The effects of the interaction are taken into account in
Sec. \ref{sec4}, where we show by an 
explicit calculation that 
the FDR's of two quadratic observables differ already at the lowest
order beyond the Gaussian approximation.
In Sec. \ref{sec5} our results are carefully compared with the
numerical and analytic ones available in the literature.
In Sec. \ref{sec6} we summarize the results obtained and their implications.
In the Appendix we report the details of the computations of the Feynman 
diagrams.

\section{The model}
\label{sec2}

One among the simplest non-trivial models displaying slow-relaxation and aging 
is a lattice spin model in $d$ dimensions with $O(N)$ symmetry and short-range
interactions evolving according to a purely dissipative dynamics 
after a quench to the critical point.  
In the simplest instance its lattice Hamiltonian is given by
\be
\H= - \sum_{\langle {\bf ij}\rangle} {\bf s_i}\cdot{\bf s_j} \;,
\label{HON}
\ee
where ${\bf s_i}$ is a $N$-component spin located at the lattice site
${\bf i}$, with ${\bf s_i}^2 = 1$. The sum runs over all 
pairs $\langle{\bf ij}\rangle$ of nearest-neighbor lattice sites. 
A purely dissipative dynamics for this model proceeds by
elementary moves that amount to random changes in the direction of the
spin ${\bf s}_{\bf i}$ (spin-flip sampling). 
The transition rates can be arbitrarily chosen provided that 
the detailed-balance condition is satisfied. 
For analytical studies the most suited are the
Glauber ones \cite{glau}, which allow exact solutions in the one-dimensional 
case \cite{lz-00,gl-00i,fins-01,mbgs-03,ms-04}.

Despite their simplicity, these models are not exactly solvable (for 
arbitrary $N$) in physical dimensions $d=2,3$, and to obtain information about
the non-equilibrium critical dynamics, one has to resort to numerical 
simulations \cite{gl-00c,bhs-01,mbgs-03,c-03,sdc-03,ak-04,c-04,ak-04b}.

To investigate analytically the dynamical behavior in physical dimensions,
we take advantage of the universality considering the time evolution
of a $N$-component field $\p({\bf x},t)$ with a
purely dissipative dynamics~(Model A of Ref.~\cite{HH}). This is
described by the stochastic Langevin equation
\be
\label{lang}
\dpar_t \p_i ({\bf x},t)=-\Omega 
\frac{\delta \cal{H}[\p]}{\delta \p_i({\bf x},t)}+\xi_i({\bf x},t) \; ,
\ee
where $\Omega$ is the kinetic coefficient, 
$\xi({\bf x},t)$ a zero-mean stochastic Gaussian noise with 
\be
\langle \xi_i({\bf x},t) \xi_j({\bf x}',t')\rangle= 2 \Omega \, \delta({\bf x}-{\bf x}') \delta (t-t')\delta_{ij},
\ee
and $\cal{H}[\p]$ is the static Hamiltonian. 
It may be assumed, near the critical point, of the Landau-Ginzburg form
\be
{\cal H}[\p] = \int d^d x \left[
\frac{1}{2} (\nabla \p )^2 + \frac{1}{2} r_0 \p^2
+\frac{1}{4!} g_0 \p^4 \right] ,\label{lgw}
\ee
where $r_0 \propto T$ is the temperature parameter, assuming its 
critical value $r_{0,c}$ for $T=T_c$ ($r_{0,c}=0$, 
within the analytical approach discussed below), and $g_0$ is the bare
coupling constant of the theory. 
This coarse-grained continuum dynamics is expected to be in the same 
universality class as the 
lattice models with $O(N)$ symmetry, short-range interactions, and 
spin-flip dynamics \cite{HH}.

The equilibrium  correlation and 
response functions can be obtained by means of the 
field-theoretical action \cite{zj,bjw-76} 
\be
S[\p,\pt]= \int d t \int d^dx 
\left[\pt \frac{\partial\p}{\partial t}+
\Omega \pt \frac{\delta \mathcal{H}[\p]}{\delta \p}-
\pt \Omega \pt\right].\label{mrsh}
\ee
Here $\pt({\bf x},t)$ is a $N$-component auxiliary field, conjugated to 
the external field $h$ in such a way that
${\cal H}[\p,h] = {\cal H}[\p] - \int d^d x h \, \p$.
As a consequence, the linear response to the field $h$ of a generic observable
${\cal O}$ is given by
\be
\frac{\delta \langle {\cal O} \rangle}{ \delta h_i({\bf x},s)} = 
\Omega \langle \tilde\p_i({\bf x},s){\cal O}\rangle \ , \quad \ i =
1,\ldots, N \ .
\label{rfield}
\ee
For this reason   $\pt({\bf x},t)$ is termed response field.

Within this field-theoretical formalism it is possible to show that the 
FDT holds for generic $R_\o$ and $C_\o$. This has an illuminating derivation
in a supersymmetric formulation, where the FDT's are the 
Ward identities due to the supersymmetry \cite{zj}. 

The effect of a macroscopic initial condition 
$\p_0({\bf x})=\p({\bf x},t=0)$ may be taken into account by 
averaging over the initial configuration
with a weight $e^{-H_0[\p_0]}$ where \cite{jss-89}
\be
H_0[\p_0]=\int\! d^d x\, \frac{\tau_0}{2}[\p_0({\bf x})-a({\bf x})]^2,
\ee
that specifies an initial state $a({\bf x})$ with Gaussian short-range 
correlations proportional to $\tau_0^{-1}$.

Following standard methods \cite{zj,bjw-76} the response and correlation 
functions may be obtained by a perturbative expansion of the functional weight
$e^{-(S[\p,\pt]+H_0[\p_0])}$ in terms of the
coupling constant $g_0$. 
The propagators~(Gaussian two-point functions of the fields $\p$ and 
$\pt$) of the resulting theory are \cite{jss-89} 
\bea
\langle \tilde{\p_i}({\bf q},s) \p_j(-{\bf q},t) \rangle_0 =& 
\delta_{ij} R^0_{\bf q}(t,s)=&\delta_{ij} \,\theta(t-s) G(t-s),\label{Rgaux}\\
\langle \p_i({\bf q},s) \p_j(-{\bf q},t) \rangle_0 =&
\delta_{ij} C^0_{\bf q}(t,s)=& \frac{\delta_{ij}}{ {\bf q}^2+r_0}\left[ G(|t-s|)+\left(\frac{r_0 +{\bf q}^2}{\tau_0}-1
\right) G(t+s)\right], \label{Cgaux}
\eea
where $\theta(t)$ is the step function [$\theta(t\le 0) = 0$,
$\theta(t>0)=1$] and 
\be
G(t)=\displaystyle{e^{-\Omega ({\bf q}^2+r_0) t}} \label{GG}.
\ee
The response function Eq. (\ref{Rgaux}) is the same as in 
equilibrium.
Eq.~(\ref{Cgaux}), instead, reduces to the equilibrium form when ${\bf
q}\neq {\bf 0}$ and both
times $t$ and $s$ go to infinity while $\tau = t - s$ is kept fixed.
In the following we will assume the \^Ito prescription 
(see, e.g., Refs.~\cite{jan-92,zj}) to deal with the 
ambiguities arising in formal manipulations of stochastic equations. 
Consequently, all the diagrams with loops of response propagators have 
to be omitted. 
This ensures that causality holds in the perturbative 
expansion~\cite{jss-89,jan-92,bjw-76}.
From the technical point of view, the breaking of time-translation invariance
does not allow the factorization of connected correlation functions in terms 
of one-particle irreducible ones as usually done when
time-translation invariance holds.
As a consequence, as it is the case when dealing with surface
critical phenomena \cite{diehl-86}, 
all the computation has to be done in terms of connected 
functions only \cite{jss-89}. Furthermore it has been 
shown \cite{jss-89} that $\tau_0^{-1}$ is an irrelevant variable for the 
RG flow affecting only the correction to the leading long-time scaling 
behavior we are interested in. In view of that 
we fix it to its fixed-point value 
$\tau_0^{-1}=0$ from the very beginning of the 
calculation.

From scaling arguments \cite{gl-00c}, and more rigorously from the solution 
of RG equations \cite{jss-89}, it is known that the zero-momentum response 
and correlation functions of the basic fields satisfy the scaling 
forms (see Ref.~\cite{cg-rev} for a review):
\bea
R_{{\bf q}=0}(t,s) &=& A_R\, (t-s)^a(t/s)^{\theta} F_R(s/t)\; ,\label{scalR}\\
C_{{\bf q}=0}(t,s) &=& A_C\,s(t-s)^a(t/s)^{\theta} F_C (s/t)\; ,
\label{scalC}
\eea
where $a = (2-\eta-z)/z$, $z$ is the dynamical
critical exponent, $\eta$ the anomalous dimension of the fields \cite{zj}, 
and $\theta$ the initial-slip exponent~\cite{jss-89,jan-92}.
We single out explicitly the non-universal amplitudes $A_{R,C}$ by fixing 
$F_{R,C}(0) = 1$. 
With this normalization $F_{R,C}$ are universal scaling functions. 
From the previous scaling forms one deduces that
\be
\dpar_s C_{{\bf q}=0}(t,s) = A_{\dpar C}\, 
(t-s)^a(t/s)^{\theta} F_{\dpar C} (s/t)\; ,
\label{scaldC}
\ee
where the non-universal amplitude $A_{\dpar C}$ has been defined so 
that $F_{\dpar C} (0) = 1$. 
Accordingly one has $A_{\dpar C} = A_C (1-\theta)$.

Using Eqs.~(\ref{scalR}) and (\ref{scaldC}) one finds that
\be
{\cal X}_{{\bf q}=0}(t,s)\equiv\frac{R_{{\bf q}=0}(t,s)}{\dpar_s C_{{\bf q}=0}(t,s)}=
\frac{A_R F_R(s/t)}{A_{\dpar C} F_{\dpar C}(s/t)}\,,
\label{Xdissut}
\ee
is a universal amplitude-ratio (in the sense of Ref. \cite{pha})
being the ratio of two quantities 
[$R_{{\bf q}=0}(t,s)$ and $\dpar_s C_{{\bf q}=0}(t,s)$] that have the same 
scaling dimensions. 
Furthermore it is a function of the ratio $s/t$ only, and not of $s$
and $t$ separately.\footnote{This is an important difference compared to
mean-field glassy model, where, instead, it turns out that $X_{{\bf x}=0}$ 
in the long-time regime 
can be written as a function of $C(t,s)$ (see, e.g., 
Refs.~\cite{cugl-02,cr-03}).}
In lattice simulations (especially in the literature concerning glassy
systems \cite{cugl-02,cr-03})
response and correlation functions are often measured in the real
space ${\bf x}$, instead of in the momentum space ${\bf q}$ as done here.
From the scaling forms given above one can derive the analogous ones in the 
real space for $R_{\bf x}$, $C_{\bf x}$, and then define 
$X_{\bf x=0}\equiv R_{\bf x=0}/\partial_s C_{\bf x = 0}$ (originally 
introduced in Refs.~\cite{ck-93,ckp-94}) in analogy with 
${\mathcal X}_{\bf q = 0}$. 
In general 
one expects $X_{\bf x = 0}\neq {\mathcal X}_{\bf q = 0}$.
Nonetheless it has been argued \cite{cg-02a1} that the
long-time limit of the universal amplitude ratio
\be
X^\infty_M=
\lim_{s\rightarrow\infty}\lim_{t\rightarrow\infty} 
\frac{R_{{\bf q}=0}(t,s)}{\dpar_s C_{{\bf q}=0}(t,s)}=
\lim_{s/t\rightarrow0}{\cal X}_{{\bf q}=0}(s/t)=
\frac{A_R}{A_{\dpar C}}=\frac{A_R}{A_C(1-\theta)}\;,
\label{Xinfamplitratio}
\ee
(the subscript $M$ refers to the fact that 
$\p_{{\bf q}=0}\propto M$, the average magnetization) is
equal to the same limit of $X_{\bf x =0}$. This equality was also 
confirmed by numerical simulations \cite{mbgs-03} (even if the numerics
of Ref.~\cite{mbgs-03} have been questioned \cite{comm}).

\subsection{Scaling forms of composite operators}

We now derive scaling forms analogous to Eqs. (\ref{scalR}) and
(\ref{scalC}) 
for the correlation and response functions 
of local composite operators, focusing on those of the form $\p^m$.
However, the derivation is completely general and can be easily applied to
any other operators.

Let us consider an observable $\o$ (a composite operator, 
using the field-theoretical terminology) having
$h_\o$ as a conjugate field (e.g., $\o$ is the energy density and $h_\o$ the 
temperature) and coupling to ${\mathcal H}$ according to
${\mathcal H} \mapsto {\mathcal H} + h_\o \o$.
As a consequence, the dynamical functional $S$
changes according to $S \mapsto S_\o=S + \Omega h_\o \ot$, 
where the associated operator $\ot$ is given by
\be
\ot=\int d t d^d x \; \pt({\bf x},t) \frac{\delta\o}{\delta\p({\bf x},t)} \;.
\label{defot}
\ee
The linear response of an observable ${\mathcal A}$ to a variation in the
field $h_\o$ can be expressed as
\be
\left .\frac{\delta \langle
{\mathcal A} \rangle_{h_\o}}{\delta h_\o}\right|_{h_\o=0} 
\equiv \Omega \langle{\mathcal A} \ot \rangle\,,
\ee
where $\langle \cdot \rangle_{h_\o}$ stands for the average over the
dynamics associated with the dynamical functional in the presence of
$h_\o$. This generalizes  Eq. (\ref{rfield}).

To render finite the correlation functions with insertions 
of the operator $\o$, of the form 
\be
\langle [\p]^n[\pt]^{\tilde n}[\o]^o[\ot]^{\tilde o}\rangle \;,
\label{gencorrfunc}
\ee
one additional renormalization (compared to those necessary when $\o$ and 
$\ot$ are not inserted, evaluated in Ref.~\cite{bjw-76})
is required: $\o_B = Z_\o \o_R$. Here and in the following with the
subscript $B$ we indicate the bare quantities and with $R$ the
renormalized ones, whose correlation functions are finite upon
removing the regularization \cite{zj}. 
In the case of operators mixing under renormalization, $\o$ has to be
understood as a vector of suitable operators, while $Z_\o$ will be in
general a renormalization matrix \cite{zj}. The presence of additive
renormalizations does not change the scaling arguments presented below. 
In view of Eq.~(\ref{defot}) one finds that $\ot_B = Z_\ot
\ot_R$ with 
\be
Z_\ot = (Z_\pt/Z_\p)^{1/2} Z_\o \;,
\label{relationZo}
\ee
where $Z_\p$ and $Z_\pt$ are the renormalization constants of the
fields, defined, as usual, 
by $\p_B = Z_\p^{1/2}\p_R$ and $\pt_B = Z_\pt^{1/2}\pt_R$.
The correlation function (\ref{gencorrfunc}) can be renormalized
according to 
\be
\langle [\p]^n[\pt]^{\tilde n}[\o]^o[\ot]^{\tilde o}\rangle_B =
Z_\p^{n/2}Z_\pt^{\tilde n/2} Z_\o^o Z_\ot^{\tilde o} \langle [\p]^n[\pt]^{\tilde n}[\o]^o[\ot]^{\tilde o}\rangle_R\,,
\ee
where, on the r.h.s., $\langle\cdot\rangle_R$ means that all the bare
quantities have been replaced by the corresponding renormalized ones.
Applying standard techniques it is possible to write the RG
equations by introducing appropriate RG functions \cite{zj}.
For the theory defined by $S_\o$, in addition to the RG functions of the 
theory with action $S$, one has to introduce
two new functions $\varrho_\o \equiv \mu\dpar_\mu \ln Z_\o|_0$ and
$\varrho_\ot \equiv \mu\dpar_\mu \ln Z_\ot|_0$\footnote{$\varrho_\o$ and 
$\varrho_\ot$ are the anomalous dimensions of $\o$ and $\ot$, respectively.
In the literature they are usually referred to as $\eta_\o$ or 
$\gamma_\o$ \cite{zj}. However, in the present case we use $\eta_\o$ and
$\gamma_\o$ to indicate critical exponents.} ($\mu$ is the scale at which the 
theory has been renormalized \cite{zj}).
With $|_0$ we indicate
that the differentiation has to be done with fixed bare parameters. 
In view of Eq.~(\ref{relationZo}), $\varrho_\o$  and $\varrho_\ot$ 
are related by 
$\varrho_\ot = \varrho_\o +(\varrho_\pt - \varrho_\p)/2$ (where
$\varrho_\p$ and $\varrho_\pt$ are the usual RG functions for the fields). 
Combining dimensional analysis with the solution of the RG equations one 
finds that the scaling dimension of the correlation function  
$\langle [\o]^o[\ot]^{\tilde o}\rangle_R$ in the 
$({\bf x},t)$-space is given, at the infrared fixed point of the theory, by
\be
\delta(o,{\tilde o}) =  o \; [\o]_\scal + \tilde o \;[\ot]_\scal\,.
\ee 
With $[\cdot]_\scal$ we 
indicate the scaling dimensions. In terms of the fixed-point values of the 
RG functions and the 
canonical (engineering) mass dimensions of the operators (denoted by
$[\cdot]_\can$) they are expressed as:
\bea
[\o]_\scal &\equiv& [\o]_\can + \varrho_\o(g^*) 
\,,\nonumber\\\phantom{}
[\ot]_\scal &\equiv& [\ot]_\can + \varrho_\ot(g^*) \,,
\eea
where $g^*$ is the fixed-point value of the renormalized coupling constant.
In analogy with the anomalous dimensions 
$\eta = \varrho_\p(g^*)$ and
$\tilde\eta = \varrho_\pt(g^*)$ (such that $[\p]_\scal =
(d-2+\eta)/2$ and $[\pt]_\scal =
(d+2+\tilde\eta)/2$ \cite{bjw-76}) one
introduces $\eta_\o$ and $\eta_\ot$:
\bea
\frac{d-2+\eta_\o}{2} &\equiv& [\o]_\scal\,,\nonumber\\
\frac{d+2+\eta_\ot}{2} &\equiv& [\ot]_\scal \ .
\eea
(Note that, in contrast to $\eta$ and 
$\tilde\eta$, $\eta_\o$ and $\eta_\ot$ do not generally vanish in the free 
theory $g^* = 0$.)
It is easy to verify that, in terms of $\eta_\o$, the critical
exponent $\gamma_\o$ of the susceptibility ($\langle \o \o \rangle
\sim |T-T_c|^{-\gamma_\o}$) is given by 
\be
\gamma_\o = \nu (2-\eta_\o) \ ,
\label{gammanu}
\ee
where $\nu$ is the critical exponent of the correlation length.
As a consequence of Eq.~(\ref{defot}) one has
$[\ot]_\can = [\o]_\can + [\pt]_\can - [\p]_\can 
= [\o]_\can + 2$ (recall that in Model A dynamics $[\p]_\can =
(d-2)/2$ and $[\pt]_\can = (d+2)/2$). 
Using $(\tilde \eta - \eta)/{2} = z-2$ (a consequence of the \fd theorem \cite{zj}), we have 
\be
\frac{\eta_\ot - \eta_\o}{2} = z - 2 
\ee
and $[\ot]_\scal = [\o]_\scal + z$, leading to
\bea
\delta(2,0) &=& d-2+\eta_\o \,,\nonumber\\
\delta(1,1) &=& d-2+\eta_\o + z\,.
\eea
Taking into account that $[{\rm time}]_\scal = -z$, one finds (hereafter 
we set $\Omega=1$)
\bea
\langle \o({\bf q},t)\o({\bf - q},s)\rangle & =& 
(t-s)^{a_\o+1}F_C({\bf q}^z(t-s),s/t)\,,
\label{scalformC}\\
\langle \o({\bf q},t)\ot({\bf - q},s)\rangle & =&
(t-s)^{a_\o}F_R({\bf q}^z(t-s),s/t)\,,
\label{scalformR}
\eea
where
\be
a_\o \equiv - \frac{\delta(1,1)-d}{z} = \frac{2 -\eta_\o - z}{z}\,.
\label{defaobs}
\ee
Let us now discuss the effect of the temporal surface.
The scaling dimension of $\pt_0$ has been computed in Ref.~\cite{jss-89}, i.e.,
$[\pt_0]_\scal = [\pt]_\scal + \eta_0/2$. 
(In terms of $\eta_0$ and $z$ the initial-slip exponent is given by 
$\theta = -\eta_0/(2 z)$ \cite{jss-89,jan-92}.)
Consider now the case of the
observables $\o^{(m)}(t)$ constructed with $m$ fields $\p$ at time
$t$, i.e. $\o^{(m)}(t)\sim \p^m(t)$. Eq.~(\ref{defot}) implies that
$\ot^{(m)}(t) \sim \pt(t)\p^{m-1}(t)$. The
observable $\widehat\o^{(m)}$, obtained from $\o^{(m)}$ by replacing
all the fields $\p$ with those $\pt$, has dimension given by
\be
[\widehat\o^{(m)}]_\scal = [\o^{(m)}]_\scal + m z\,.
\label{dimhat}
\ee
Moreover, keeping in mind the scaling dimension of $\pt_0$  
one has, for $t=0$,
\be
[\widehat\o^{(m)}_0]_\scal = [\widehat\o^{(m)}]_\scal - m\theta z\,,
\label{dimhatzero}
\ee
where $\widehat\o^{(m)}_0 = \widehat\o^{(m)}(t=0)$.

When inserted into correlation functions, 
$(\dpar_t \p)_{t=0} = 2 \Omega \pt_0$, while the insertion of an initial 
field $\p_0$ vanishes \cite{diehl-86}. 
Thus, for the  operators 
$\o^{(m)}$ and $\ot^{(m)}$ we expect 
the following short-distance expansion for $t\rightarrow 0$
\bea
\o^{(m)}(t) & \sim& \rho(t) \pt_0^m + \mbox{h.o.c.f.}\,, \\
\ot^{(m)}(t)& \sim& \tilde\rho(t) \pt_0^m + \mbox{h.o.c.f.}\,,
\eea
where \mbox{h.o.c.f.} stands for higher order composite fields that
contribute to this expansion only with subleading terms.
As a consequence, 
the scaling dimensions of $\rho$ and $\tilde\rho$ are given by
\be
[\rho]_\scal 
= - m z + m \theta z
\quad \mbox{ and } \quad
[\tilde{\rho}]_\scal 
=- (m -1)z +m \theta z,
\ee
where we used Eqs.~(\ref{dimhat}), (\ref{dimhatzero}), and the FDT. 
Taking into account that $[{\rm time}]_\scal = -z$, one 
concludes that, for $s\rightarrow 0$,
\be
\o^{(m)}(s) \sim s^{m - m \theta}\\
\quad \mbox{ and } \quad
\ot^{(m)}(s)\sim s^{m-1-m \theta}\;.
\ee
It is now possible to rewrite the scaling
forms~(\ref{scalformC}) and~(\ref{scalformR}) in a way that shows
explicitly the behavior of the scaling functions $F_C$ and $F_R$ for
$s\rightarrow 0$, i.e.,
\bea
\langle \o^{(m)}({\bf q},t)\o^{(m)}({\bf - q},s)\rangle & =& s
(t-s)^{a_\o}(t/s)^{-(m-1)+m\theta} \hat F^\o_C({\bf q}^z(t-s),s/t)\,,
\label{scalformCfin}\\
\langle \o^{(m)}({\bf q},t)\ot^{(m)}({\bf - q},s)\rangle & =&
(t-s)^{a_\o}(t/s)^{-(m-1)+m\theta} \hat F^\o_R({\bf q}^z(t-s),s/t)\,,
\label{scalformRfin}
\eea
where now the functions $\hat F^\o_C$ and $\hat F^\o_R$ are regular for
$s\rightarrow 0$. Furthermore they are also universal once we fix the 
normalization for small arguments.

Let us focus on the scaling properties of the correlation and response
functions of the observables with $m=2$. Because of the $O(N)$ symmetry
of the underling theory only two quadratic zero-momentum 
operators with different scaling dimensions exist,
that can be written as \cite{zj}
\bea
E(t) &\equiv& \int \!d^d x\; \sum_{i=1}^N \p_i^2({\bf x},t) = 
\int \!
\frac{d^d q}{(2\pi)^d}\; \sum_{i=1}^N  \p_i({\bf q},t)\p_i({\bf -q},t)\,, \label{defE}\\
T_{ij}(t) &\equiv& \int \!d^d x\; \p_i({\bf x},t)\p_j({\bf x},t)-\frac{1}{N}\delta_{ij}E(t)= 
\int \!\frac{d^d q}{(2\pi)^d}\; \p_i({\bf q},t)\p_j({\bf -q},t) -\frac{1}{N}\delta_{ij}E(t)\;.
\label{defT}
\eea
The corresponding response operators are given by
\bea
\widetilde E(t) &=& \sum_{i=1}^N \int \!
 \frac{d^d q}{(2\pi)^d}\; 2 \pt_i({\bf q},t)\p_i({\bf -q},t)\;,\\
\widetilde T_{ij}(t) &=& 
\int \!\frac{d^d q}{(2\pi)^d} \; [\pt_i({\bf q},t)\p_j({\bf -q},t) +
\p_i({\bf q},t)\pt_j({\bf -q},t)]- \frac{1}{N}\delta_{ij}\widetilde E(t)\;.
\eea
In the following we will refer to them, generically, as $\o(t)$.
The corresponding scaling functions are
\bea
C^\o(t,s) &\equiv& \langle \o(t)\o(s)\rangle = A^\o_C\,s
(t-s)^{a_\o}(t/s)^{-1+2\theta} F^\o_C(s/t)\;,
\label{scalCOfin}\\
R^\o(t,s) &\equiv& \langle \o(t)\ot(s)\rangle =
A^\o_R \,(t-s)^{a_\o}(t/s)^{-1+2\theta} F^\o_R(s/t)\;,
\label{scalROfin}
\eea
where the non-universal amplitudes $A^\o_{C,R}$ are determined so 
that $F^\o_{C,R}(0) =1$.

In terms of these quantities we can write the (zero-momentum) FDR as 
\be
{\mathcal X}^\o(t,s) \equiv \frac{\Omega R^\o(t,s)}{\dpar_s C^\o(t,s)}\,,
\ee
(here we restore the actual value of $\Omega$, currently set to $1$)
that, as its analogous (\ref{Xdissut}), depends only on the ratio $s/t$
and it is a universal function. 
In particular its universal long-time limit reads
\be
{\mathcal X}^\infty_\o = 
\lim_{s\rightarrow \infty}\lim_{t\rightarrow \infty} {\mathcal X}_\o(t,s)=
\frac{1}{2}\frac{ A^\o_R}{(1-\theta)A^\o_C}  \;.
\label{AAA}
\ee

\subsubsection{Overview of the known values of the exponents}

For the two specific cases of the quadratic operators $E(t)$ and $T_{ij}(t)$, 
the exponent $a_\o$ can be expressed in terms of more familiar 
critical exponents. 
In fact we obtained $a_\o=(2-\eta_\o-z)/z$, with $\eta_\o$ scaling
dimension of the operator $\o$. Using Eq.~(\ref{gammanu}) 
we can express it in terms of the 
susceptibility exponent, given 
by\footnote{When hyperscaling holds (i.e., for $d\leq 4$) $2-d\nu$ can be 
replaced by $\alpha$, the specific heat exponent.} (see, e.g., 
Ref.~\cite{cpv-02})
\be
\gamma_\o = 
\left\{ 
\begin{array}{cc}
2-d\nu\,,
&\quad \o = E\,, \\ 
-d\nu+2\phi_T\,,
& \quad \o = T\,,
\end{array}
\right.
\end{equation}
where $\phi_T$ is the 
so-called quadratic crossover exponent \cite{zj,cpv-02}.
The exponents $\gamma_E$ and $\gamma_T$ are exactly known in $d=2$ \cite{zj}.
Even in $d=3$ they are known with very high accuracy, in fact they
have been computed up to five loops in the $\e$-expansion 
\cite{5loop} and six (sometimes seven) loops in fixed dimension $d=3$
\cite{6loop,cpv-02} for generic values of $N$ (see Ref.~\cite{PV-r} 
for a review).
The dynamic critical exponents are currently known with a much less
accuracy. In fact $z$ has been computed up to three loops in the 
$\e$-expansion \cite{av-84}
and up to four in fixed dimensions \cite{zrev}, whereas 
$\theta$ in known only up to two loops in the $\e$-expansion \cite{jss-89}. 
For $d>4$ the mean-field results hold: 
$a_\o=1-d/2$ (for both $E$ and $T$) and $\theta=0$.

For later convenience, we report here explicitly the 
$O(\e)$ expansion of $a_\o$\footnote{In passing, let us mention that, 
for $N=1$, $\phi_T$ has a
non-trivial value, even if the operator $T_{ij}(t)$ is not defined
for the Ising model. This fact has an interpretation in terms of a gas of 
$N-$color interacting loops (see, e.g., Ref.~\cite{n-82}) 
belonging to the same universality class as the $O(N)$ model.} 
(recall that $z=2+O(\e^2)$ \cite{HH})
\begin{equation}
a_\o = 
\left\{ 
\begin{array}{cc}
{\displaystyle -1 + \frac{4-N}{2(N+8)} \epsilon + O(\epsilon^2)}\;,  &\quad \o
= E \\ \\
{\displaystyle -1 + \frac{N+4}{2(N+8)} \epsilon + O(\epsilon^2)}\;, &
\quad \o
= T
\end{array}
\right.
\label{avalues}
\end{equation}
and $\theta$
\be
\theta = \frac{N+2}{N+8}\frac{\e}{4} + O(\e^2)\;.
\label{thvalue}
\ee

The exponents are exactly known in the limit $N\rightarrow\infty$:
$\nu^{-1}=d-2,\,z=2,\, \theta=1-d/4$ \cite{jss-89}, 
and $\phi_T=2\nu$ in $d<4$, whereas
$\nu=1/2,\,z=2,\,\theta=0$, and $\phi_T=1$ in $d>4$ (mean-field exponents).
Using these values, the scaling forms for the energy in $d<4$ are
\bea
C^E(t,s) & =& A^E_C\,s
(t-s)^{d/2-3}(t/s)^{1-d/2} F^E_C(s/t)\;,\label{scalformC2mdm4}\\
R^E(t,s) &= & A^E_R \,(t-s)^{d/2-3}(t/s)^{1-d/2} F^E_R(s/t)\;,
\label{scalformR2mdm4}
\eea
whereas for  $T$ in $d<4$,
\bea
C^T(t,s) & = A^T_C\,s
(t-s)^{1-d/2}(t/s)^{1-d/2} F^T_C(s/t)\;,\label{scalformC3md}\\
R^T(t,s) &=
A^T_R \,(t-s)^{1-d/2}(t/s)^{1-d/2} F^T_R(s/t) \;.
\label{scalformR3md}
\eea
For $d>4$ the scaling forms for $E$ and $T$ are the same:
\bea
C^\o(t,s) & = A^\o_C\,s
(t-s)^{1-d/2}(s/t) F^\o_C(s/t)\;,\label{scalformC4md}\\
R^\o(t,s) &=
A^\o_R \,(t-s)^{1-d/2}(s/t) F^\o_R(s/t) \;.
\label{scalformR4md}
\eea

\section{The Gaussian approximation}
\label{sec3}

For the Gaussian model the response and correlation 
functions are known exactly, so we can evaluate the FDR (in Ref.~\cite{ckp-94} 
this has been done directly in real space).
From Eqs.~(\ref{Rgaux}), (\ref{Cgaux}), and the definition (\ref{Xdissut}),
properly generalized to ${\bf q}\neq0$, one finds \cite{cg-02a1}
\be
\displaystyle
{\cal X}_{\bf q}^0(t,s)=
\frac{\Omega R^0_{\bf q}}{\dpar_s C^0_{\bf q}} =
\frac{1}{1+e^{-2 \Omega ({\bf q}^2+r_0) s}}.
\ee
If the theory is non-critical ($r_0 \neq 0$) the limit of this ratio for 
$s\rightarrow \infty$ is $1$ for all the values of ${\bf q}$, in 
agreement with 
the idea that in the high-temperature phase all the fluctuating modes have a 
finite equilibration time, so that equilibrium is recovered and the
FDT applies.
In the critical theory 
the limit ratio is again equal to one when ${\bf q}\neq 0$, whereas
for ${\bf q}=0$ one has
${\cal X}_{{\bf q}=0}^0(t,s)=1/2$. 
This shows that the only mode that ``does not relax'' to the 
equilibrium is the zero mode in the critical limit. 
This picture (already presented in Ref.~\cite{cg-02a1}) 
has been confirmed by a one-loop computation \cite{cg-02a1}.

In the case of the Gaussian theory, with $g_0 = 0$ in the Hamiltonian 
(\ref{lgw}), the FDR can be easily computed for a generic
observable. In particular we
show that the FDR for a set of local one-point observables  
is always equal to $1/2$ in the long-time limit.
This allows for a definition of a unique effective temperature in the 
Gaussian model.
We consider 
operators of the form $\o_{i,n}=\dpar^i\p^n$. The correlation and response
functions of all local operators can be written in terms of those 
of $\o_{i,n}$.

\begin{figure}[t]
\centerline{\epsfig{file=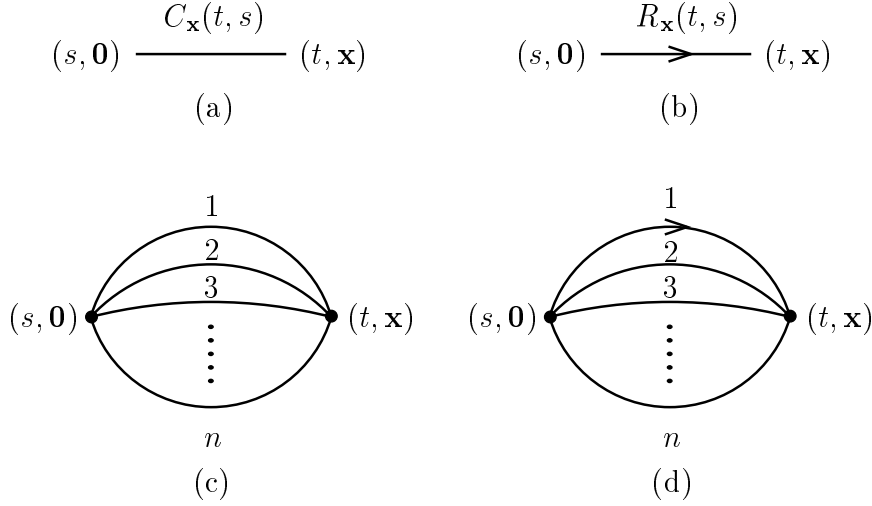}} 
\caption{%
Diagrammatic elements: (a) correlation and (b) response
functions of the order parameter. 
(c) Diagrammatic representation for the two-point correlation function 
$C^\o_{\bf x}(t,s) = \langle \o_{0,n}(t,{\bf x})
\o_{0,n}(s,{\bf 0})\rangle$ and 
(d) response function  
$R^\o_{\bf x}(t,s) = \langle \o_{0,n}(t,{\bf x}) \widetilde{\o}_{0,n}(s,{\bf
0})\rangle$. See the text for further explanations.
}
\label{diagram}
\end{figure}

The critical (i.e., $r_0=0$) response and correlation functions of the order 
parameter Eqs.~(\ref{Rgaux}) and (\ref{Cgaux}) are given, in real space, by
(their diagrammatic representation is reported in Fig. \ref{diagram} (b)
and (a), respectively)
\bea
R_{\bf x}(t,s)&=& F_d(t-s,{\bf x})\,,\\
C_{\bf x}(t,s)&=& K_{d}(t-s,{\bf x})-K_{d}(t+s,{\bf x})\,,
\eea
where (we set $\Omega=1$) 
$F_d(\tau,{\bf x})=\int (dq) e^{-{\bf q}^2 \tau}e^{-i {\bf q}\cdot {\bf x}}$
and 
$K_d(\tau,{\bf x})=\int (dq) {\bf q}^{-2} e^{-{\bf q}^2 \tau}e^{-i {\bf q}\cdot {\bf x}}$ (with $({\rm d}q) = {\rm d}^dq/(2\pi)^d$).

Let us explain our argument considering first the operators $\o_{0,n}$. 
The two-point correlation functions of $\o_{0,n}$ is given by the
$(n-1)-$loop diagram with the two points connected by $n$ correlation lines, 
as shown in Fig.~\ref{diagram} (c).

In the real space its expression is simply given by 
the product of $n$ correlators.
Thus
\be
C^\o_{\bf x}(t,s)=c_n [C_{\bf x}(t,s)]^n\,,
\ee
($c_n$ is the combinatorical factor associated with the diagram) 
whose derivative is
\be
\dpar_s C^\o_{\bf x}(t,s)=c_n n 
[C_{\bf x}(t,s)]^{n-1} \dpar_s C_{\bf x}(t,s)\,.
\ee
Analogously the response function is given by the diagram depicted in
Fig.~\ref{diagram} (d), obtained from that one contributing to the
correlation function (Fig.~\ref{diagram} (c)) 
by replacing an order-parameter correlator with a response function:
\be
R^\o_{\bf x}(t,s)=c_n n[C_{\bf x}(t,s)]^{n-1} R_{\bf x}(t,s)\,,
\ee
where the factor $n$ comes from the fact that the response operator is 
$\widetilde{\o}_{0,n}=n \o_{0,n-1} \pt$. Note that the 
combinatorial factor $c_n$ is the same as for the correlation function.

The FDR is given by
\be
X_{\bf x}^{\o_{0,n}}(t,s)\equiv\frac{R^{\o_{0,n}}_{\bf x}(t,s)}{\dpar_s C^{\o_{0,n}}_{\bf x}(t,s)}=
\frac{R_{\bf x}(t,s)}{\dpar_s C_{\bf x}(t,s)}\equiv X^M_{\bf x}(t,s)\,.
\label{relalltimes}
\ee
Thus we obtain the remarkable result that the FDR of {\it powers of the field}
in real space is equal, for all times, to the FDR of the field.

Before considering the effects of the derivatives let us consider the
previous relation in momentum space. 
Remembering that the product of two functions
after a Fourier transformation becomes a convolution, one has 
\bea
R_{\bf q}^\o(t,s)           &=&c_n n (C * \cdots *C 
*R)_{\bf q} \,,\\
\dpar_s C_{\bf q}^\o(t,s)&=& c_n n (C * \cdots *C 
*C')_{\bf q}\,,
\eea 
where $*$ is the convolution, $C$ and $R$ are in momentum
space (with the time dependence understood), $\cdots$ means $n-1$ times, and 
$C'_{\bf q}=\dpar_s C_{\bf q}$. For ${\bf q}=0$ the previous relations become
\bea
R_{{\bf q}=0}^\o(t,s)     &=&c_n n \int (dp) (C * \cdots *C)_{\bf p} R_{-{\bf p}} \,,\\
\dpar_s C_{{\bf q}=0}^\o(t,s)&=&c_n n \int (dp) (C * \cdots *C)_{\bf p} C'_{-{\bf p}}\,.
\eea
Thus 
\be
\left[{\mathcal X}_{{\bf q}=0}^\o(t,s)\right]^{-1}=
\frac{\int (dp) (C * \cdots *C)_{\bf p} C'_{-{\bf p}}} { \int (dp) (C
* \cdots *C)_{\bf p} R_{-{\bf p}} } =
\frac{\int (dp) (C * \cdots *C)_{\bf p} R_{-{\bf p}} {\cal X}^{-1}_{-{\bf p}}}{ \int (dp) (C * \cdots *C)_{\bf p} R_{-{\bf p}} }\,,
\label{XX}
\ee
i.e., the inverse of ${\mathcal X}_{{\bf q}=0}^\o$ is a weighted average of 
${\cal X}^{-1}_{-{\bf p}}$, 
with weight $(C * \cdots * C)_{\bf p} R_{-{\bf p}}$, that in the limit 
$t\rightarrow\infty$, $s\rightarrow\infty$ in the proper order, is
peaked around ${\bf p} = 0$, giving 
${\mathcal X}^\infty_\o\equiv 
\lim_{s\rightarrow\infty} \lim_{t\rightarrow\infty} 
{\mathcal X}^\o_{{\bf q}=0}(t,s)=
\lim_{s\rightarrow\infty} \lim_{t\rightarrow\infty} {\mathcal X}_{{\bf
p}=0}(t,s) = X_M^\infty$. 
Note that in momentum space, at variance with the relation 
Eq.~(\ref{relalltimes}) holding
between FDR in real space, only the 
long-time limit of the FDR reproduces the FDR for the fields. 
This is in agreement with some explicit calculations we 
made for $\p^2$ (see below) and $\p^3$ FDR's.\footnote{The Gaussian $\p^3$ FDR 
is obtained from Eqs. (A9) and (B6) of Ref. \cite{cg-02a2}. The calculation
is straightforward, but cumbersome.}

Let us now take into account the effect of the derivatives. In momentum space 
they amount simply to 
a multiplication by ${\bf q}^{i}$ and so they change the 
weight in Eq. (\ref{XX}) by a factor ${\bf q}^{2i}$ (one ${\bf q}^i$ for 
each insertion), not affecting our conclusion on the long-time limit.

Let us note that if instead we introduce some non-local operators, as e.g. 
$e^{\dpar^2 \o}$ (with $\o$ a local operator), the weight in 
Eq. (\ref{XX}) is no longer exponential in ${\bf q}^2$ and the long-time 
limit of FDR may be different from $1/2$.

\begin{figure*}
\begin{center}
\epsfig{file=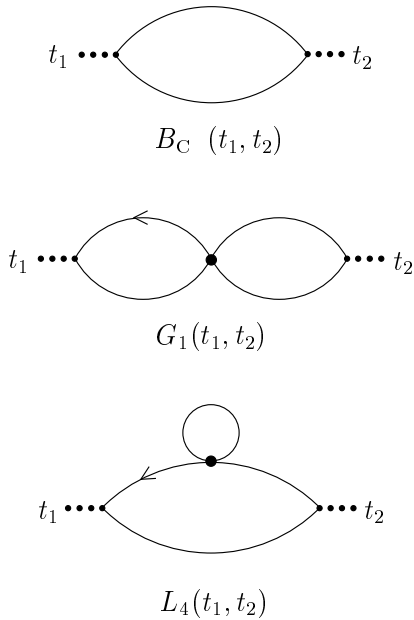,width=0.33\textwidth} 
\end{center}
\caption{Diagrams contributing to the correlation function.
$G_1(t,s)$ and $L_4(t,s)$ have also symmetric counterparts that have been 
considered.}
\label{diagC}
\end{figure*}

\section{Two-loop computation}
\label{sec4}

Here we present the details of the two-loop (i.e., up to $O(\e)$ in the 
$\e$-expansion) perturbative computation of the
correlation and response functions for the zero-momentum energy and the 
tensor operators given by Eqs. (\ref{defE}) and (\ref{defT}).
The one-loop results are the Gaussian ones obtained in the previous section.

\begin{figure*}
\begin{center}
\epsfig{file=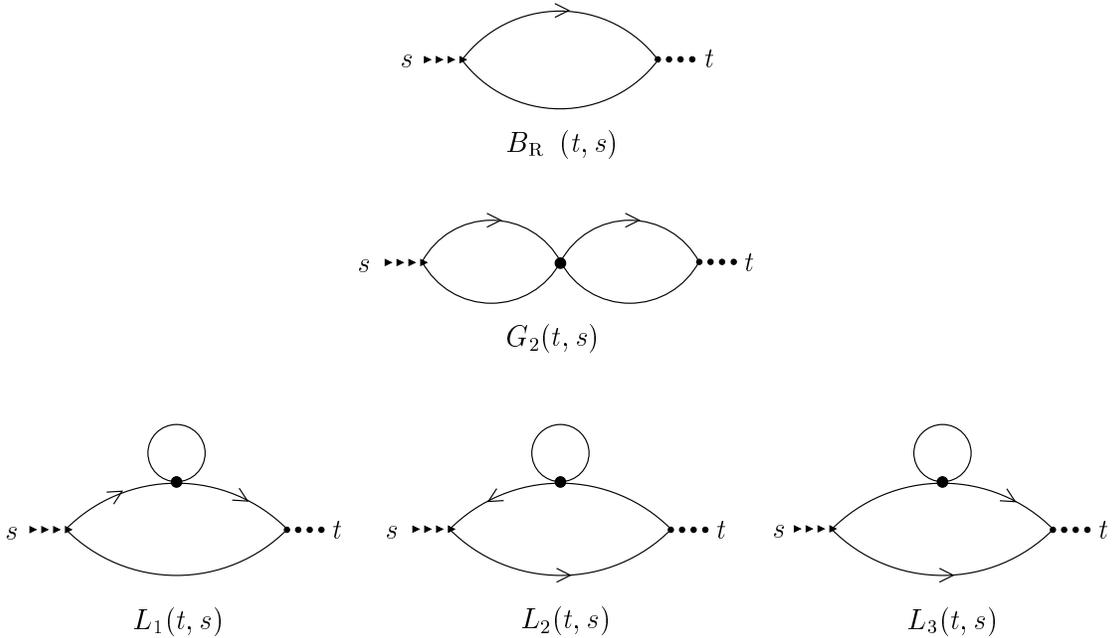,width=0.9\textwidth} 
\end{center}
\caption{Diagrams contributing to the response function.}
\label{diagR}
\end{figure*}

The diagrammatic contributions are reported in Fig.~\ref{diagC} for
the correlation function $C^\o(t,s)$ and in Fig.~\ref{diagR} for the
response function $R^\o(t,s)$. In Table~\ref{Tab:coeff} we list the
symmetry factors of the diagrams (depending both on the global
topology and on the external legs) and the corresponding 
color factors (depending on global topology and the index structure
of the vertices).

For later convenience let us introduce the function
\begin{equation}
I_n(t) = \int\!\!\frac{d^dq}{(2\pi)^d}\frac{e^{-2q^2t}}{q^{2n}} =
I_n(1) t^{-d/2+n}\,,
\label{defI}
\end{equation}
where
\begin{equation}
I_n(1) = N_d 2^{n-d/2-1} \Gamma(d/2-n)\,,
\end{equation}
and $N_d = 2/[(4\pi)^{d/2}\Gamma(d/2)]$.

Thus we obtain, for the one-loop 
diagrams\footnote{Note that without the temporal surface at $t=0$,
i.e., using the equilibrium correlator $C^{(\rm eq)}(t_1-t_2)$,
$B_{\rm C}$ would be given only by $I_2(|t_1-t_2|)$. 
For $d = 4-\e$, $I_2$ has a pole [$I_2(1) \propto \Gamma(2-d/2)$], 
that is usually subtracted by introducing the well-known additive 
renormalization for the $\p^2$ correlation function. 
In this case, however, the time-boundary term yields two additional factors 
that exactly cancel the pole of the first term. 
Accordingly, in contrast to the equilibrium case, no additive
renormalization has to be introduced to render finite the energy
correlation function. 
Thus the equilibrium {\it renormalized} correlation functions for ${\bf q}=0$
cannot be simply
recovered by taking the limit $t,s\rightarrow\infty$ with $t-s$ fixed, 
as usually done for field correlation functions.}
\bea
B_{\rm C}(t_1,t_2) &=& \int\!\!\frac{d^dq}{(2\pi)^d} [C^0_{\bf q}(t_1,t_2)]^2 =
I_2(|t_1-t_2|) - 2 I_2(\max\{t_1,t_2\}) + I_2(t_1+t_2)\,,
\label{bollaCC}\\
B_{\rm R}(t,s)&=& \int\!\!\frac{d^dq}{(2\pi)^d} C^0_{\bf q}(t,s) R^0_{\bf q}(t,s) = 
I_1(t-s) - I_1(t)\,,
\eea
where here and hereafter we set $\Omega = 1$ to lighten the notation and 
$t>s$ (for $t<s$ the diagrammatic contributions are zero due to the 
causality of the response functions).

\begin{table}[t]
\caption{Symmetry and color factors of the Feynman diagrams 
depicted in Figs. \ref{diagC} and \ref{diagR}.}
\begin{tabular}{lcc|rcc}
&\multicolumn{2}{c|}{Symmetry} & \multicolumn{3}{c}{Color}\\
& Corr. & Resp. & coeff. & $E$ & $T$ \\
\hline
$B_{\rm R},B_{\rm C}$ & $2$ & $2$ & $C_1=$ & $N$ & $1/2$ \\
$G_1,G_2$ & $2$ & $2$ & $C_{2a}=$& $N(N+2)/3$ & $1/3$ \\
$L_1,L_2,L_3,L_4$ & $2$ & $1$ & $C_{2b}=$ & $N(N+2)/3$ & $(N+2)/6$ \\
\end{tabular}
\label{Tab:coeff}
\end{table}

In terms of the two-loop Feynman integrals the response and correlation 
functions read 
\bea
R^\o_{\rm B}(t,s) &=& 4 C_1 B_{\rm R}(t,s) -  4 C_{2a} g G_2(t,s) -   
2 C_{2b} g [L_1(t,s)+L_2(t,s) +L_3(t,s)] + O(g^2)\,,\label{sumR}\\
C^\o_{\rm B}(t,s) &=& 2 C_1 B_{\rm C}(t,s) - 2 C_{2a} 
g [G_1(t,s)+G_1(s,t)]-  2 C_{2b} g [L_4(t,s) + L_4(s,t)] + O(g^2)\label{sumC}\,.
\eea
The calculations of $G_i$ and $L_i$  are quite cumbersome and all
the details are reported in the Appendix. 
Taking into account the expressions therein we find 
renormalized response and correlation functions that are in agreement with 
equations (\ref{scalROfin}) and (\ref{scalCOfin}), with the exponents given in 
Eqs. (\ref{avalues}) and (\ref{thvalue}).
The non-universal amplitudes and scaling functions are given by 
(as usual we introduce $\tilde g \equiv N_d g$ to simplify the notation)
\bea
A_{R}^{E,T} & =& \left[ 1-\frac{\e}{2}+ \left(\frac{\e}{2}
-\frac{C_{2a}}{C_1}\frac{\tilde g}{2} \right)(\gamma_E + \log 2) + \frac{C_{2b}}{C_1}\frac{\tilde
g}{4}\right] N_d C_1 \,,
\label{AR}\\
F_{R}^{E,T}(x) & =& 1 + \left( \frac{\e}{2} - \frac{C_{2b}}{C_1}\frac{\tilde
g}{4}\right)\left[\frac{\log (1-x)}{x} + 1 - \log(1-x) \right] \,,
\label{FR}\\
A_{ C}^{E,T} &=& \left[ 1-\frac{\e}{2}+ \left(\frac{\e}{2}
-\frac{C_{2a}}{C_1}\frac{\tilde g}{2} \right)(\gamma_E + \log 2) + 
\frac{C_{2b}}{C_1}\frac{\tilde g}{2} + 
\frac{C_{2a}}{C_1}\frac{5}{24}\tilde g\right]N_d C_1 \,,
\label{AC}\\
F_{C}^{E,T}(x) &=&\frac{1-x}{x^2}\left[ -\log(1-x^2) 
-\frac{\e}{4} W_1(x) -\frac{C_{2a}}{C_1}\frac{\tilde g}{4} W_2(x)
- \frac{C_{2b}}{C_1}\frac{\tilde g}{2} W_3(x)\right]\,,
\label{FC}
\eea
where the functions $W_i(x)$ have been defined in Eqs.~(\ref{W1}), (\ref{W2}),
and (\ref{W3}). The previous equations are valid up to 
$O(\tilde g^2,\e\tilde g,\e^2)$.

According to  Eq.~(\ref{AAA}), 
${\mathcal X}^\infty_\o$ can be expressed as an amplitude ratio
\be
{\mathcal X}^\infty_\o = \frac{A_R^\o}{2A_C^\o(1-\theta)} \,.
\ee
Using the non-universal constants given in Eqs.~(\ref{AR})
and~(\ref{AC}) we find that
\begin{equation}
{\mathcal X}^\infty_\o = \frac{1}{2}\left(1-\frac{C_{2b}}{C_1}\frac{\tilde
g^*}{8} - \frac{C_{2a}}{C_1} \frac{5}{24}\tilde g^* \right) + O(\e^2)\,,
\end{equation}
where 
\be
\tilde g^* = \frac{6}{N+8} \e + O(\e^2)\,,
\ee
is the fixed-point value of the coupling constant \cite{zj}.
Taking into account the combinatorical factors in Table \ref{Tab:coeff} we 
finally find:
\begin{equation}
{\mathcal X}_\o^\infty = \left\{
\begin{array}{cc}
{\displaystyle \frac{1}{2}\left(1-\frac{2}{3}\frac{N+2}{N+8}\e\right) +
O(\e^2)}\;,&\quad \o = E\;,\\ \\
{\displaystyle \frac{1}{2}\left(1-\frac{1}{12}\frac{3N+16}{N+8}\e\right)
+ O(\e^2)}\;, &\quad \o = T \;.
\end{array}
\right.
\label{XinfO}
\end{equation}
Recall that the one-loop FDR for the order parameter (magnetization) is
\cite{cg-02a1}
\begin{equation}
{X}_M^\infty = \frac{1}{2}\left(1-\frac{1}{4}\frac{N+2}{N+8}\e\right)
+ O(\e^2)\,.
\label{MFDR}
\end{equation}
From these results we conclude that the long-time limit of the 
fluctuation-dissipation ratio depends on the particular observable
chosen to compute it.

As a final remark let us comment on the connection between the correlation
functions of $\p^2$ in Model A dynamics and those of the conserved density 
$\en$ in the associated Model C \cite{HH}. 
In equilibrium dynamics it is usually easier to compute 
$\langle \p^2\p^2\rangle$ in terms of $\langle\en\en\rangle$ (see, e.g., 
Ref.~\cite{fm-03}).
However, this is no longer possible when studying non-equilibrium dynamics
(considered in Ref.~\cite{cg-03}) because of the connection between 
$\p$ and $\en$ does not carry over to this case.

\section{Comparison with other results}
\label{sec5}

\subsection{The spherical model and the $N\rightarrow\infty$ limit}

The static critical behavior of the $O(N)$ model in the limit 
$N\rightarrow\infty$ is known to be equivalent to that of the spherical
model, defined by the Hamiltonian
\be
\H=\frac{1}{2}\sum_{\langle {\bf ij}\rangle}(s_{\bf i}-s_{\bf j})^2\,,
\ee
where $s_{\bf i}$ are real numbers 
subjected to the constraint $\sum_{\bf i}s_{\bf i}^2=L^d$ ($L$ being the 
linear dimension of the $d$-dimensional lattice, assumed for simplicity to 
be hypercubic), and the sum runs over all 
nearest-neighbor pairs $\langle {\bf ij}\rangle$.
Indeed Stanley proved \cite{stanley-68} that the free energies of the 
two models are exactly the same. From this equality it 
follows that critical exponents, scaling functions etc. are equal.
The same equivalence
also holds for equilibrium critical dynamics defined in the spherical model  
by means of a Langevin equation as Eq. (\ref{lang})\footnote{%
This equation has to be properly 
modified in order to prescribe a dynamics that is
compatible with the spherical constraint.%
}.
As far as we are aware, this equivalence has not yet been carried over to the 
non-equilibrium critical dynamics we are interested in.
Let us recall that the free energies of the 
models in the presence of a spatial boundary (whose corresponding field theory
looks very similar to the non-equilibrium one that we are considering) 
are well known to be not equal \cite{sfbound}.

The spherical model has attracted a lot of attentions, since its essential 
Gaussian Hamiltonian makes it exactly solvable although the resulting 
critical behavior is not mean-field like because of the spherical constraint.
The FDR of the magnetization was calculated by Godr\`eche and Luck, who 
obtained $X_M^\infty=1-2/d$. 
This result is compatible with the two-loop 
$\e$-expansion \cite{cg-02a2} for $N=\infty$ and, moreover, it agrees 
with the exact result we are going to derive. 
Recently Sollich \cite{s04} has determined several FDR's of 
quadratic (in the spin $s_{\bf i}$) operators that could be 
compared with our results. He considered the bond energy observable 
$B_{\bf i}=\frac{1}{2}(s_{\bf i}-s_{\bf j})^2$, the product observable 
$P_{\bf i}=s_{\bf i} s_{\bf j}$ (with ${\bf i},{\bf j}$ 
nearest neighbors)  both in the real and in the momentum space, 
and the total energy. The {\it exact} results of his analysis are that
$X^\infty_P=X^\infty_B=X^\infty_M\neq X^\infty_E$ (the expansion of 
$X^\infty_E$ close to four dimensions for $d<4$ nicely agrees with our 
two-loop $\e$-expansion for $N=\infty$). 
The scaling forms for these observables have been also derived and they agree
with our equations (\ref{scalformR}) and (\ref{scalformC}) with 
$a_B=-1-d/2$, $a_P=1-d/2$, and $a_E=d/2-3$ in $d<4$ (and 
$a_E=a_P=a_B+2=1-d/2$ in $d>4$). 
All these findings agree with our calculation apart from the fact that 
the observable $P$ of the spherical model 
cannot be naively identified with $\p^2$ for $N=\infty$.
On the other hand $a_B=a_P-2$ agrees with the naive identification of $B$ as 
the Laplacian of $P$ in the continuum limit.

%%%%%%%%%%%%%5

We report now the calculation of some FDR's for the $O(N)$ model 
directly for $N=\infty$,
using the well-known property that for $N=\infty$ the fourth order 
interaction term can be self-consistently decoupled 
$g/N (\p^2)^2\rightarrow g C_{{\bf x}=0}(t) \p^2$ (here 
$g_0$ is replaced by $g/N$) \cite{zj}. 
Taking advantage of this decoupling the 
exact response function of the theory (for $2<d<4$) has
been computed \cite{jss-89}, finding:
\be
R_{\bf q}(t,s)=\theta(t-s) \left(\frac{t}{s}\right)^{1-d/4}e^{-{\bf q}^2(t-s)}
\,,\label{RN}
\ee 
whereas, for the correlation function \cite{jss-89}
\be
C_{\bf q}(t,s)=2\int_0^\infty dt' R_{\bf q}(t,t') R_{\bf q}(s,t')=
2(t s)^{1-d/4} e^{-{\bf q}^2(t+s)}\int_0^s dt'\, t'^{d/2-2}e^{2{\bf q}^2t'}\,.
\label{CN}
\ee
In particular for ${\bf q}={ 0}$  the previous expressions become
\be
R_{{\bf q}={ 0}}(t,s)=\theta(t-s) \left(\frac{t}{s}\right)^{1-d/4}\quad
\mbox{and}\quad
C_{{\bf q}={ 0}}(t,s)=\frac{2}{d/2-1} s  \left(\frac{t}{s}\right)^{1-d/4}\,.
\ee
According to Eq. (\ref{Xinfamplitratio}) it is straighfroward to compute the 
FDR for the order parameter, finding $X_M^\infty = 1-2/d$, the same result 
as for the spherical model.

\begin{figure}[t]
\centerline{\epsfig{width=13truecm,file=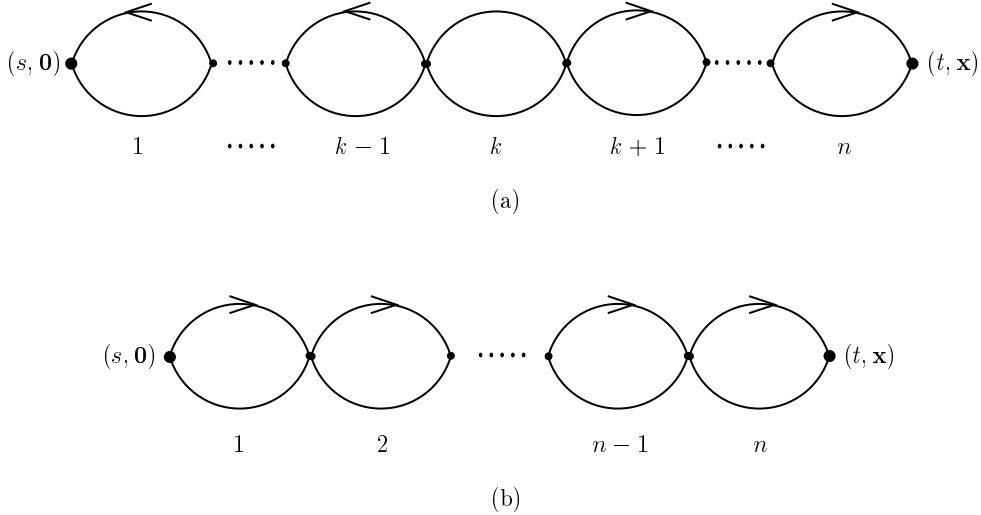}} 
\caption{%
Chains of bubble diagrams for the (a) correlation 
and (b) response function of quadratic observables.
For the correlation (a), the index $k$ runs between
$1$ and $n$.  
}
\label{chain}
\end{figure}

Now we  
consider, in the limit $N\rightarrow\infty$,
the scaling functions (\ref{scalCOfin}) and (\ref{scalROfin})
of the response and the correlation 
of the quadratic operators.
The explicit computation for $2<d<4$ of the scaling functions,
the non-universal amplitudes, and, eventually, the
FDR of composite operators is not as straightforward as one could 
have erroneously expected.
In fact, if one roughly assumes, on the sole basis of the decoupling
in the large-$N$ limit, that the theory is essentially Gaussian with 
renormalized two-point functions given by Eqs. (\ref{RN}) and (\ref{CN}), 
the result $X^\infty_\o=X^\infty_M$ would follow from the argument we gave
for the Gaussian Model, that makes no use of the specific expressions of the
response and correlation functions.

Let us consider more closely the case of the energy $E$.
For the two-point function $\langle E(t) E(s)\rangle$, a 
family of diagrams (chains of bubble diagrams, see Fig. \ref{chain}) 
that are of order $O(N^0)$ even in the limit 
$N\rightarrow\infty$ exists (although the coupling constant is of order $1/N$,
each bubble carries a combinatorical factor $N$). These diagrams are not
accounted for by a simple 
renormalization of the two-point functions of the order parameter. 
This fact is well known for static observables (see, e.g., the 
calculation of the structure factors 
$\langle\p^2({\bf q}) \p^2({\bf -q})\rangle$ 
up to four loops in Ref. \cite{cpv-02}).
If and only if the contribution of such diagrams to $X^\infty$ 
vanishes (or is cancelled out) one has
$X^\infty_E=X^\infty_M$. The results of the previous sections indicate that 
this is not the case.

Note that if one considers instead of $E$ the operator $T$ then
the chains of bubble diagrams are depressed by a combinatorical 
factor of order $1/N$ and so they do not contribute to the correlation
and response functions for $N=\infty$. 
On this sole basis one concludes that in the limit $N=\infty$,
$X^\infty_T=X^\infty_M$ to all order in $\e$ and not only at 
the first one as we have explicitly obtained (compare Eq. (\ref{XinfO})
with Eq. (\ref{MFDR}) considering the limit $N\rightarrow\infty$).

%%%%%%%%%%%%%%%%%%

This analysis  indicates that the observable $P$ in the spherical model 
has a scaling behavior (and $X^\infty$) that is the same as that of $T$, 
contrarily to the naive expectation, suggesting instead $E$.

Finally let us mention that for all the quantities we considered here, the
non-equilibrium dynamics of the spherical and $O(\infty)$ models are exactly
the same.  
This fact calls for a more rigorous investigation of a
possible correspondence between the two models beyond the case of
equilibrium dynamics.

\begin{figure}[t]
\centerline{\epsfig{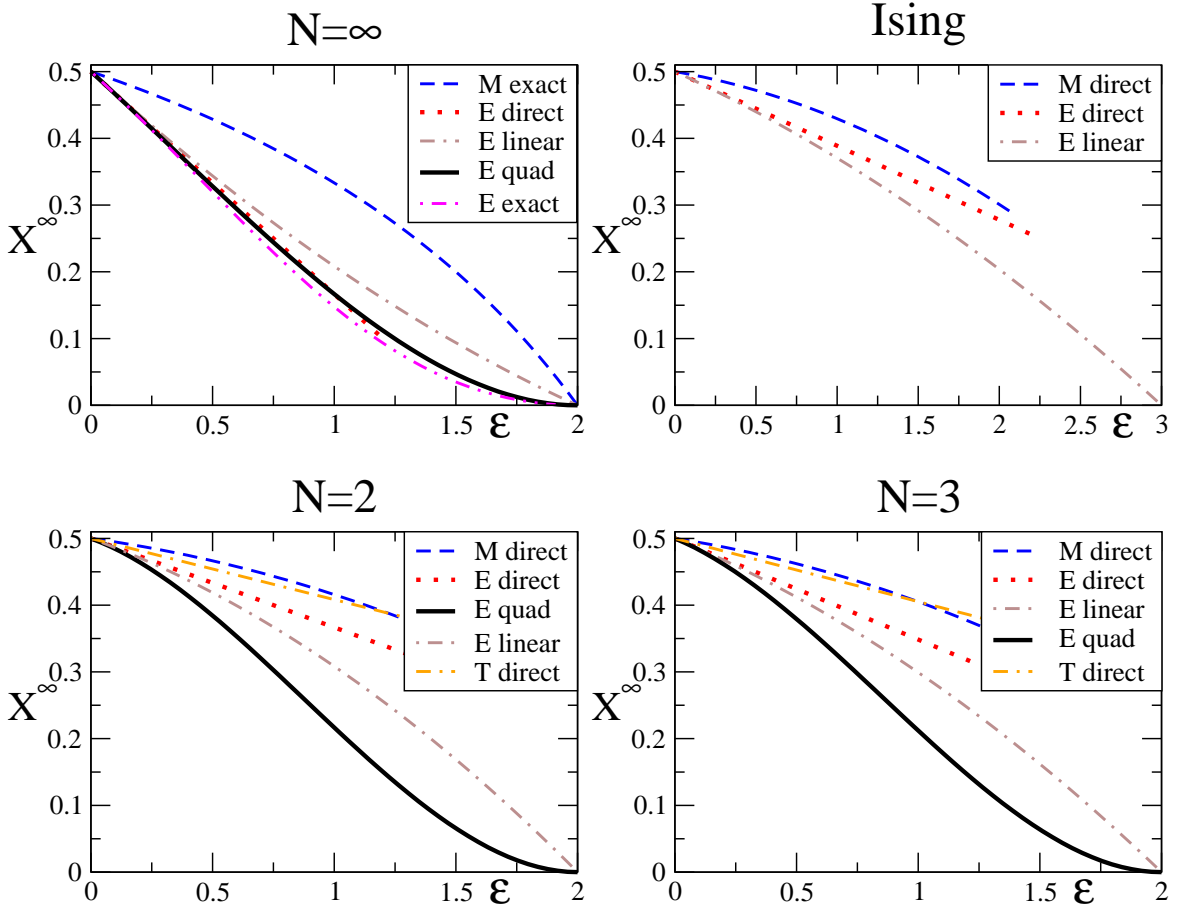}\vspace{1.3cm}} 
\caption{%
$X^\infty_\o$ for $\o=M,E,T$ and $N=1$,$2$,$3$, and $\infty$.
For $X^\infty_M$ we report the two-loop result \protect\cite{cg-02a2}.
For $X^\infty_E$ we report the direct estimate and those obtained by 
constraining at the lower critical dimensions 
(with linear constraint for $N=1,2,3,\infty$ and with quadratic one 
for $N=2,3,\infty$).
For $X^\infty_T$ only the direct estimate is reported.
For $N=\infty$ we also report as ``E exact'' 
the exact result for the spherical model, from Ref.~\protect\cite{s04}.
}
\label{XETM}
\end{figure}

\subsection{The Ising model}

Apart from the spherical model, the FDR of composite operators has been 
considered so far only for the one- and two-dimensional Ising models,
analytically \cite{mbgs-03,ms-04} and numerically \cite{mbgs-03}.

In one dimension the several FDR's considered \cite{mbgs-03,ms-04} 
turned out to equal $X^\infty_M=1/2$, apart from that of the energy: 
$X_E^\infty=0$. 
This fact has been interpreted in Ref. \cite{mbgs-03}
as an interplay between criticality and coarsening,
a peculiarity of those models (such as the one-dimensional Ising model)
with $T_c=0$. Instead, 
our result indicates that $X^\infty_E<X^\infty_M$ is a 
more general property, rigorously true close to four dimensions ($\e\ll1$)
and perhaps valid up to $\e=3$.

In two dimensions, the numerical results 
$X^\infty_M=0.340(5)\simeq X^\infty_E=0.33(2)$ apparently indicate the 
equality of the two FDR's \cite{mbgs-03}. 
Obviously our result calls for a more precise determination of 
$X^\infty_E=0.33(2)$ to understand whether this apparent equality is 
due to the relatively low precision of such a measure or to the fact that 
$d=2$ is a
peculiar case for some still unknown reasons.
In any case, let us stress that $X^\infty_E$ from the $O(\e)$
result for $\e=2$ gives a value much smaller than $X^\infty_M$.
The direct estimate (unconstrained [u]) from the two-loop series in Eq. 
(\ref{XinfO}) for 
$\e=2$ (giving $X^\infty_E{\rm [u]} \sim 0.28$) is 
probably unreliable as it was the case for $X^\infty_M$.  
($X^\infty_M{\rm [1loop]}\sim 0.42$ \cite{cg-02a1} whereas
$X^\infty_M{\rm [2loop]}=0.30(5)$ \cite{cg-02a2}.) 
To obtain a more reliable result (without computing the $O(\e^2)$ term,
that seems to be a very difficult calculation, requiring 
the evaluation of three-loop 
diagrams) one can constrain linearly [l] the $O(\e)$ 
result to assume the exactly known value for $d=1$ (i.e., $\e=3$), 
as usually done for this kind of expression (see, e.g., Ref.~\cite{PV-r}).
Assuming a smooth behavior in $\e$ up to $\e=3$, one can write
\be
X^\infty_E{\rm [l]} =\frac{1}{2}\left(1-\frac{\e}{3}\right)\left[1 + \frac{\e}{9}+O(\e^2)\right]\,,
\label{constr}
\ee 
that has the same $\e$-expansion as Eq. (\ref{XinfO}), but it is expected
to converge more rapidly to the correct result. From
Eq. (\ref{constr}) we get for the two-dimensional Ising model $X^\infty_E{\rm [l]}\sim 0.20$, 
that is much lower than 
the value that has been determined so far
$X^\infty_M\simeq 0.33$ \cite{cg-02a2,c-03,mbgs-03,sdc-03,c-04}.
However, we stress that a robust field-theoretical prediction for 
$X^\infty_E$ for the two-dimensional Ising model requires a (difficult)
higher-loop computation.
For the three-dimensional Ising model we obtain
$X^\infty_E{\rm [l]}\simeq0.37$, to be compared with the direct estimate 
$X^\infty_E{\rm [u]}\simeq0.39$. 
Note that, as usual in $d=3$, the spreading of the different estimates
is much smaller, signaling a higher reliability of these predictions. 
(We recall that, in three dimensions, the field-theoretical estimate
for
the FDR of the magnetization is 
$X^\infty_M{\rm [2loop]}=0.429(6)$ \cite{cg-02a2}.)

In Fig. \ref{XETM} we report Eq. (\ref{constr}) for $0<\e<3$ compared with 
the direct estimate Eq. (\ref{XinfO}) and with the two-loop FDR of the 
magnetization \cite{cg-02a2}.

\subsection{The $O(N)$ model with $N\geq2$.}

Eq. (\ref{XinfO}) allows us to provide predictions for the $O(N)$ models
with arbitrary $N$. So far the non-equilibrium dynamics of models with
continuous symmetry ($N>1$) 
has been numerically studied only for the $XY$ ($N=2$) model, both in  
$d=2$ \cite{bhs-01,ak-04} and $d=3$ \cite{ak-04b}. 
The value that has been determined in $d=3$, i.e., 
$X^\infty_M=0.43(4)$ \cite{ak-04b} is 
in good agreement with the two-loop field-theoretical
prediction $X^\infty_M=0.416(8)$ \cite{cg-02a2}. 
The observable dependence of $X^\infty_\o$ has not yet been addressed
in these cases.

To obtain an estimate of $X^\infty_{E,T}$ for the three-dimensional $O(N)$
model we again constrain the $\e$-expansion
at the lower-critical dimension ($d_{{\rm l.c.d.}}=2$ in this case). 
We assume $X^\infty_{E}(d=2)=0$ for $N\geq 2$. This surely holds for $N>2$,
since for $d=2$ these systems are in the coarsening regime. 
Such an assumption is instead
questionable for $N=2$, where it is known that a finite-temperature phase 
transition of topological nature takes place at $T_{KT}$. 
A linear constraint would lead to
\be
X^\infty_E{\rm[l]}=\frac{1}{2}\left(1-\frac{\e}{2}\right)\left[1 +\frac{16-N}{6(N+8)} \e+O(\e^2)\right]\,.
\label{lincon}
\ee 
On the other hand the exact results 
for the spherical model \cite{s04} suggest that the approach 
to $d=2$ is quadratic rather than linear, so it is tempting to
implement a quadratic constraint also for finite $N$
\be
X^\infty_E{\rm[q]}=\frac{1}{2}\left(1-\frac{\e}{2}\right)^2 \left[1 +\frac{20+N}{3(N+8)}\e+O(\e^2)\right]
\label{quadcon}\,.
\ee 
In Fig. \ref{XETM}, we report Eqs. (\ref{lincon}) and (\ref{quadcon}) 
for $0<\e<2$ and $N=2,3,\infty$. 
It is apparent that for $N=\infty$ the 
approximation obtained implementing the quadratic constraint 
reproduces better the result for the spherical model.
For the three-dimensional $XY$ model ($N=2$) we 
get $X^\infty_E\simeq 0.37$ from direct 
estimate (no constraint, [u]), $X^\infty_E\simeq 0.31$ from linear
constraint [l], 
and $X^\infty_E\simeq 0.22$ from the quadratic one [q].
For the three-dimensional Heisenberg model ($N=3$) 
we find, instead, $X^\infty_E{\rm [u]}\simeq 0.35$, 
$X^\infty_E{\rm [l]}\simeq 0.30$, and 
$X^\infty_E{\rm [q]}\simeq 0.21$.

Even the results with constraints are rather scattered, making very difficult 
to provide robust estimates in $d=3$. However, we can surely conclude that 
$X^\infty_E<X^\infty_M$ for all $2<d<4$ and that their difference should be
large enough to be observed in Monte Carlo simulations in three dimensions.
The analysis of the non-equilibrium behavior within the $\tilde{\e}=d-2$
expansion \cite{zj} may clarify which, between the linear and the
quadratic constraint, is the proper one close to $d=2$, 
even if the results for
the spherical model strongly suggest the latter.

For $X^\infty_T$ we note that it is very close to $X^\infty_M$, even for 
$\e\ll 1$ (see Fig. \ref{XETM}), making the numerical detection of such
a difference probably very difficult.

\section{Conclusions}
\label{sec6}

In this paper we considered the problem of the definition of a {\it unique} 
effective temperature via the long-time limit of the 
\fd relation for critical systems quenched from 
a high-temperature phase to the critical point and evolving according to 
a purely dissipative dynamics. 
Within the field-theoretical approach to non-equilibrium 
critical dynamics and by means of appropriate RG equations, we 
obtained the general scaling forms for the response and correlation functions
of a generic local observable $\o(t)$, from which it is possible to 
derive the FDR $X_\o(t,s)$.

We found that in the Gaussian approximation all the local 
operators have $X^\infty_\o=1/2$, allowing for a definition of a unique 
effective temperature. This equality is broken already at the 
first order in the $\e$ expansion for the quadratic operators we considered 
(namely the total energy and the tensor, see Eqs. (\ref{defE}) 
and (\ref{defT})).
Let us point out that our results go further than those obtained for the
one-dimensional Ising model \cite{mbgs-03,ms-04} and for the spherical 
model \cite{s04}. 
In these cases $X^\infty_\o = X^\infty_M$ for all the observables $\o$, 
except for the total energy. This operator is conjugated to the temperature 
of the bath but not to the actual one (if any) of the system. 
On this basis one could doubt that the energy operator is not as suited as 
others to define the effective temperature, resulting in a different 
$X^\infty$. Nevertheless we find that there is at least one further operator,
namely $T_{ij}$, having $X_T^\infty\neq X^\infty_M, X^\infty_E$.
This explicitly shows that, at variance with what is often conjectured,
a unique effective temperature can not be defined for this 
class of models.

Our results for $N\rightarrow\infty$ are always in agreement with the recent
ones for the spherical model \cite{s04}, calling for a proof of a
possible correspondence (or for a counterexample).
They instead disagree (at qualitative level) with the available 
numerical simulation
of the Ising model \cite{mbgs-03} apparently giving $X^\infty_E=X^\infty_M$.
Probably a more accurate measure of $X^\infty_E$ is required to detect the 
difference (if any) between $X^\infty_E$ and $X^\infty_M$.
We also provided theoretical predictions for the $O(N)$ model for arbitrary
$N$ in $2<d<4$. It should be possible to check them quantitatively in 
$d=3$, where the $\e$-expansion is expected to be more accurate.
This calls for numerical simulations or real experiments in three-dimensional 
systems.

Let us finally comment that it would be interesting to understand how the 
standard scenario of effective temperature \cite{ckp-97,cugl-99,cugl-02} 
can be generalized to the case when each sector of a theory has a different 
$T_{\rm eff}^\o$, as our results indicate to be the case for critical
systems beyond the mean-field approximation.

\section*{Acknowledgments}

We thank L. Berthier, H.~W. Diehl,  A. Pelissetto, M. Pleimling, and 
P. Sollich for stimulating discussions, correspondence, and suggestions. 
In particular we are indebted to Peter Sollich for providing us the 
unpublished work \cite{s04} and for a patient and careful comparison 
between our results, obtained with very different methods.
We thank L. Cugliandolo and S. Franz for partially inspiring this work
and ICTP for hospitality.
AG acknowledges the Max-Planck Institut f\"ur Physik komplexer Systeme
(MPIPKS) in Dresden for warm hospitality during the NESPHY03 Seminar,
where part of the computation presented here has been done.
PC acknowledges financial support from EPSRC Grant No. GR/R83712/01.

\appendix

\section{Two-loop Feynman diagrams}
\label{app}

In this appendix we report all the details of the evaluation of two-loop
integrals.
Here and in the following we will denote $t_< = \min\{t_1,t_2\}$ and
$t_> = \max\{t_1,t_2\}$.
For the diagram $G_1$ one finds
\bea
G_1(t_1,t_2) &=& \int_0^{t_1} \!\!d t' B_{\rm R} (t_1,t') B_{\rm
CC}(t',t_2) \nonumber\\
&=& \int_0^{t_1} \!\!d t' I_2(t'+t_2) [I_1(t_1-t') - I_1(t_1)] \nonumber\\
&& + \int_0^{t_<} \!\!d t' [I_2(t_2-t') - 
2 I_2(t_2)] [I_1(t_1-t') - I_1(t_1)]\nonumber\\
&& + \theta(t_1-t_2) \int_{t_2}^{t_1} \!\!d t' [I_2(t'-t_2) - 
2 I_2(t')] [I_1(t_1-t') - I_1(t_1)] \,.
\eea
To isolate the dimensional poles of this expression one has to keep in
mind that $I_1(t)$ diverges for $t\rightarrow 0$ and $d=4$. These
singularities have to be removed by proper subtractions. The remaining
part can be expanded in a regular power series of $\e = 4 -d$. 
Then one finds that
\bea
G_1(t_1,t_2) &=& I_2(1)I_1(1) t_1^\e\bigg\{ \int_0^1\!\!d x
\left[\left(x+\frac{t_2}{t_1}\right)^{\e/2} \!\!\! - \left(1+\frac{t_2}{t_1}\right)^{\e/2}\right][(1-x)^{-1+\e/2}-1] \nonumber\\
&&+\left(\frac{2}{\e}-1\right)\left(1+\frac{t_2}{t_1}\right)^{\e/2}\nonumber\\
&&+ \theta(t_1-t_2) \int_0^{t_2/t_1}\!\!d x \left[\left(\frac{t_2}{t_1} - x\right)^{\e/2}\!\!\! -
2\left(\frac{t_2}{t_1}\right)^{\e/2}\right][(1-x)^{-1+\e/2}-1] \nonumber\\
&&+ \theta(t_1-t_2) \int_{t_2/t_1}^1\!\!d x \left[\left(x-\frac{t_2}{t_1}\right)^{\e/2} - 2x^{\e/2}
-\left(1 - \frac{t_2}{t_1}\right)^{\e/2} + 2\right][(1-x)^{-1+\e/2}-1]
\nonumber\\
&&+\theta(t_1-t_2)\left[ \left(
1-\frac{t_2}{t_1}\right)^{\e/2}-2\right]\left[ \frac{2}{\e}\left(
1-\frac{t_2}{t_1}\right)^{\e/2} - \left(1-\frac{t_2}{t_1}\right)\right]\nonumber\\
&&+ \theta(t_2-t_1) \int_0^1\!\!d x\left[ \left(
\frac{t_2}{t_1}-x\right)^{\e/2} -
\left(\frac{t_2}{t_1}-1\right)^{\e/2}\right][(1-x)^{-1+\e/2}-1]\nonumber\\
&&+ \theta(t_2-t_1) \left(\frac{2}{\e}-1\right)\left[\left(\frac{t_2}{t_1}-1\right)^{\e/2}\!\!\! - 2
\left(\frac{t_2}{t_1}\right)^{\e/2}\right]\bigg\}\,.
\eea
Expanding this expression in power of $\e$, one gets
\bea
G_1(t_1,t_2) &=&
-\frac{N_d^2}{4}\frac{1}{\e}\log\left(1-\frac{t_<^2}{t_>^2}\right)
%\nonumber\\&&
-\frac{N_d^2}{4}(\gamma_E + \log 2 -\frac{1}{2} + \log
t_>)\log\left(1-\frac{t_<^2}{t_>^2}\right)\nonumber\\
&&+\frac{N_d^2}{8}\frac{t_2}{t_1}\left[
\log\left(1+\frac{t_<}{t_>}\right) -
\log\left(1-\frac{t_<}{t_>}\right)\right]\nonumber\\
&&-\frac{N_d^2}{8}\left[\log\frac{t_<}{t_>}\log\left(1-\frac{t_<}{t_>}\right)+\frac{3}{2}\log^2\left(1+\frac{t_<}{t_>}\right)+\frac{1}{2}\log^2\left(1-\frac{t_<}{t_>}\right)\right]\nonumber\\
&&-\frac{N_d^2}{8}\theta(t_1-t_2)
\bigg[2\frac{t_2}{t_1}-\frac{\pi^2}{3} +
\log^2\left(1-\frac{t_2}{t_1}\right) -
\log\frac{t_2}{t_1}\log\left(1+\frac{t_2}{t_1}\right)\nonumber\\
&&\hspace{2.5cm} +2\li\left(1-\frac{t_2}{t_1}\right) +
\li\left(-\frac{t_2/t_1}{1-t_2/t_1}\right) +
\li\left(\frac{t_2/t_1}{1+t_2/t_1}\right)\bigg]\nonumber\\
&&-\frac{N_d^2}{8}\theta(t_2-t_1)
\bigg[2 -\frac{\pi^2}{6} - \li\left(-\frac{t_1/t_2}{1-t_1/t_2}\right)
+ \li\left(\frac{1}{1+t_1/t_2}\right)\bigg] + O(\e)\,.
\eea
For the following computations it is useful to introduce the function
$W(x)$ defined as
\bea
W(x) &=& -2 +\frac{\pi^2}{2} -2x + 2\log(1-x^2) + 
\left(x+\frac{1}{x}\right)\log\frac{1+x}{1-x} \nonumber\\
&& - 2 \log x \log(1-x) + \log x
\log(1+x) - 2 \log^2(1-x) \nonumber\\
&& - 3\log^2(1+x) - 2 \li(1-x) - \li\left(
\frac{1}{1+x}\right) - \li\left(\frac{x}{1+x}\right)\,,
\label{defW}
\eea
whose expansion for small $x$ is $W(x) = -17 x^2/6 + O(x^3)$. $W(x)$
enters the expression of $G_1(t,s) + G_1(s,t)$ as
\bea
G_1(t,s) &+& G_1(s,t) = \nonumber\\&&
-\frac{N_d^2}{2}\frac{1}{\e}\log\left(1-\frac{s^2}{t^2}\right) -
\frac{N_d^2}{2}(\gamma_E + \log 2 + \log
t)\log\left(1-\frac{s^2}{t^2}\right) + \frac{N_d^2}{8}W(s/t) \;.
\eea

The expression of the diagram $G_2$ is
\begin{equation}
G_2(t,s) =  \int_s^t \!\!d t' B_{\rm R}(t,t') B_{\rm R}(t',s) \,,
\end{equation}
after the proper subtractions one finds
\bea
G_2(t,s) &=&I_1^2(1)t^{-1+\e}\bigg\{ \int_{s/t}^1\!\!d x
\left[(1-x)^{-1+\frac{\e}{2}}-\left(1-\frac{s}{t}\right)^{-1+\frac{\e}{2}}\right]
%\nonumber\\&&\hspace{2.5cm}\times
\left[\left(x-\frac{s}{t}\right)^{-1+\frac{\e}{2}}
-\left(1-\frac{s}{t}\right)^{-1+\frac{\e}{2}}\right]
 \nonumber\\
&&+\left(\frac{4}{\e}-1\right)\left(1-\frac{s}{t}\right)^{-1+\e} 
- \int_{s/t}^1\!\!d x (1-x)^{-1+\e/2}(x^{-1+\e/2}-1) \nonumber\\
&&-\frac{4}{\e}\left(1-\frac{s}{t}\right)^{\e/2} +
 \frac{2}{\e}\left[1-\left(\frac{s}{t}\right)^{\e/2}\right]\bigg\}\,,
\eea
whose expansion is given by
\be
G_2(t,s) = \frac{N_d^2}{4}\frac{s}{t}\frac{1}{t-s}\bigg\{
\frac{1}{\e} + \gamma_E + \log 2 + \log t + \frac{1}{2}\left(
1+\frac{t}{s}\right)\log \left(1-\frac{s}{t}\right)\bigg\} + O(\e)\,.
\ee
In order to compute the diagrams $L_i$ one has to determine the tadpole
(loop of the correlation function), given by
\begin{equation}
P(t) = \int\!\!\frac{d^dq}{(2\pi)^d} C^0_{\bf q}(t,t) = - I_1(t)\,,
\end{equation}
where the dimensional regularization has been used. The contribution
of the tadpole to the two-point correlation function of the order
parameter (i.e., $\langle \p({\bf
q},t_1)\p({\bf -q},t_2)\rangle$, a subdiagram of $L_i$), is given by
\bea
D_{\rm RC}(t_1,t_2;q) &=& \int_0^{t_1}\!\!d t'
R^0_{\bf q}(t_1,t')P(t')C^0_{\bf q}(t',t_2) \nonumber\\
&=& - \left[ \int_0^{t_<}\!\!d t' I_1(t')
\frac{e^{-q^2(t_1+t_2-2t')}}{q^2} -
I_1(1)\frac{t_<^{2-d/2}}{2-d/2}\frac{e^{-q^2(t_1+t_2)}}{q^2}\right]\nonumber\\
&&-\theta(t_1-t_2)I_1(1) C^0_{\bf q}(t_1,t_2)\frac{t_1^{2-d/2}-t_2^{2-d/2}}{2-d/2}\,.
\eea
While the contribution to the response function $\langle \p({\bf
q},t)\pt({\bf -q},s)\rangle$ is given by
\be
D_{\rm RR}(t,s;q) =  \int_s^t\!\!d t'
R^0_{\bf q}(t,t')P(t')R^0_{\bf q}(t',s) 
= -I_1(1) \frac{t^{2-d/2}-s^{2-d/2}}{2-d/2} R_q(t,s)\,.
\ee
Using the previous expressions, the diagram $L_1$ can be written as
\bea
L_1(t,s) &=&  \int\!\!\frac{d^dq}{(2\pi)^d}C^0_{\bf q}(t,s)D_{\rm RR}(t,s;q)
= -I_1(1) \frac{t^{2-d/2}-s^{2-d/2}}{2-d/2} B_{\rm R}(t,s) 
\nonumber\\&&
= -I_1^2(1) \frac{t^{2-d/2}-s^{2-d/2}}{2-d/2}[(t-s)^{1-d/2}-t^{2-d/2}]\,,
\label{eqL1}
\eea
whose expansion is
\be
L_1(t,s) = \frac{N_d^2}{16} \frac{s}{t}\frac{1}{t-s}\log\frac{s}{t} + O(\e)\,.
\ee
Analogously
\bea
L_2(t,s) =  \int\!\!\frac{d^dq}{(2\pi)^d}R^0_{\bf q}(t,s)D_{\rm RC}(s,t;q)
= -\int_0^s\!\!d t' I_1(t')I_1(t-t') + I_1(1)I_1(t)
\frac{s^{2-d/2}}{2-d/2} \,,
\label{eqL2}
\eea
and subtracting from the integrand its singular behavior for
$t'\rightarrow 0$, we find
\be
L_2(t,s) = -\int_0^s\!\!d t' I_1(t')[I_1(t-t')-I_1(t)]
= -I_1^2(1) t^{3-d}\int_0^{s/t}\!\!d x x^{1-d/2}[(1-x)^{1-d/2}-1]\,,
\ee
whose expansion is
\be
L_2(t,s) = \frac{N_d^2}{16}\frac{1}{t}\log\left(1-\frac{s}{t}\right) +O(\e)\,.
\ee
The expression for $L_3$ is given by
\bea
L_3(t,s)&=& \int\!\frac{d^dq}{(2\pi)^d}
R^0_{\bf q}(t,s)D_{\rm RC}(t,s;q) \nonumber\\
&=& - \int_0^s\!\!d t' I_1(t') I_1(t-t') + 
I_1(1)  I_1(t) \frac{s^{2-d/2}}{2-d/2}  -
I_1(1)\frac{t^{2-d/2}-s^{2-d/2}}{2-d/2} B_{\rm R}(t,s) \,,
\eea
that, keeping into account Eqs.~(\ref{eqL1}) and~(\ref{eqL2}), leads to
\bea
L_3(t,s) = L_2(t,s) + L_1(t,s)\,.
\eea
The last diagram is $L_4$, for which we find
\bea
L_4(t_1,t_2) &= & \int\!\!\frac{d^dq}{(2\pi)^d}C^0_{\bf q}(t_1,t_2)D_{\rm
RC}(t_1,t_2;q)\nonumber\\
&= & 
- \bigg\{\int_0^{t_<}\!\!d t' I_1(t')
[I_2(t_>-t') - I_2(t_1+t_2-t')] 
\nonumber\\&&\hspace{2.5cm} 
- I_1(1)\frac{t_<^{2-d/2}}{2-d/2}[I_2(t_>)-I_2(t_1+t_2)]\bigg\}
\nonumber\\&& 
- \theta(t_1-t_2)I_1(1) \frac{t_1^{2-d/2}-t_2^{2-d/2}}{2-d/2} B_{\rm C}(t_1,t_2)\,.
\eea
Subtracting the singular part of the integrands one finds
\bea
L_4(t_1,t_2) &= & -I_1(1)I_2(1) \nonumber\\
&& \times \bigg\{
 t_>^\e \int_0^{t_</t_>}\!\!d x x^{-1+\e/2}\left[(1-x)^{\e/2} -
\left(1+\frac{t_<}{t_>} - x\right)^{\e/2} - 1 +
\left(1+\frac{t_<}{t_>}\right)^{\e/2}\right]\nonumber\\
&& + \theta(t_1-t_2) \frac{2}{\e}\left( t_1^{\e/2} -
t_2^{\e/2}\right) \left[ (t_1-t_2)^{\e/2} +
(t_1+t_2)^{\e/2} - 2 t_1^{\e/2}\right]\bigg\}\,,
\eea
whose expansion is
\be
L_4(t_1,t_2) =  \frac{N_d^2}{8} \left[ \li\left(
\frac{t_</t_>}{1+t_</t_>}\right)-
\li\left(\frac{t_<}{t_>}\right)\right]- \theta(t_1-t_2)
\frac{N_d^2}{8} \log\frac{t_2}{t_1} \log\left( 1 -
\frac{t_2^2}{t_1^2}\right) + O(\e)\,.
\ee

%%%%%%%%%%%%%%%%%%%%%%%%%%%%%%%%%%%%

Inserting the previous expressions into Eqs. (\ref{sumR}) and 
(\ref{sumC}) we end up for the response function with
\bea
R_{\rm B}(t,s) &=& 
\left(1-\frac{C_{2a}}{C_1}\frac{\tilde g}{\e}\right) 
A_R \frac{s}{t}\frac{1}{t-s}
\nonumber\\&&\times
\left[1+\left(\frac{\e}{2} - \frac{C_{2a}}{C_1}\frac{\tilde
g}{2}\right) \log(t-s) \right]\left[1+ \frac{C_{2b}}{C_1}\frac{\tilde
g}{4}\log \frac{t}{s}\right] F_R(s/t) + O(\tilde g^2,\e\tilde
g,\e^2 )\,,
\label{bareRhat}
\eea
and for the correlation function
\bea
C_{\rm B}(t,s) &=& 
\left(1-\frac{C_{2a}}{C_1}\frac{\tilde g}{\e}\right) 
A_{C} \frac{s^2}{t} \frac{1}{t-s}
\nonumber\\&&\times
\left[1+\left(\frac{\e}{2} - \frac{C_{2a}}{C_1}\frac{\tilde
g}{2}\right) \log(t-s) \right]\left[1+ \frac{C_{2b}}{C_1}\frac{\tilde
g}{4}\log \frac{t}{s}\right] F_{C}(s/t) + O(\tilde g^2,\e\tilde
g,\e^2 )\,,
\label{bareChat}
\eea
where $A_R$, $F_R$, $A_C$, and $F_C$ are the expressions reported in the text
[Eqs. (\ref{AR}), (\ref{FR}), (\ref{AC}), and (\ref{FC})].
To shorten the formulae we introduce the following functions:
\bea
W_1(x)&=& \log^2(1+x) - \log^2(1-x)-2\log(1-x)\log(1+x)
+ 2 \log(1-x^2)\,, \label{W1}\\
W_2(x)&=&W(x) + \log^2(1+x) + 3
\log^2(1-x) + 2 \log(1-x)\log(1+x)-\frac{5}{6}\log(1-x^2)\,, \label{W2}\\
W_3(x)&=& \li\left(\frac{x}{1+x}\right) - \li(x) -
\log(1-x^2) \label{W3}\,.
\eea

The bare expressions~(\ref{bareRhat}) and~(\ref{bareChat}) 
have to be renormalized according to [using the fact that $\o_B =
Z_\o \o_R$, $\ot_B = Z_\ot\ot_R$,
$\Omega_B = (Z_\p/Z_\pt)^{1/2}\Omega_R=\Omega_R+O({\tilde
g}^2,\e\tilde g, \e^2)$, and 
$Z_\ot=(Z_\pt/Z_\p)^{1/2} Z_\o = Z_\o + O({\tilde g}^2, {\tilde g}\e,\e^2) =\left(1-\frac{C_{2a}}{2C_1}\frac{\tilde
g}{\e}\right)+O({\tilde g}^2,{\tilde g}\e, \e^2)$ \cite{zj}]
\bea
C_R(t,s) &= \left(1+\frac{C_{2a}}{C_1}\frac{\tilde g}{\e}\right)  C_B(t,s)\,, \nonumber\\
R_R(t,s) &= \left(1+\frac{C_{2a}}{C_1}\frac{\tilde g}{\e}\right)  R_B(t,s)\,. 
\eea
Exponentiating the logarithms, we recover the expected scaling forms and 
exponents with 
\be
a_\o+1=\frac{\e}{2} -\frac{C_{2a}}{C_1}\frac{\tilde g^*}{2} = 
\left\{ 
\begin{array}{cc}
E: & \quad {\displaystyle \frac{4-N}{2(N+8)}}\e+O(\e^2) \,,\\
\\
T: & \quad {\displaystyle \frac{N+4}{2(N+8)}}\e+O(\e^2) \,,
\end{array}
\right.
\ee
and
\begin{equation}
2\theta=\frac{C_{2b}}{C_1}\frac{\tilde g^*}{4} = \frac{N+2}{N+8}
\frac{\e}{2}+O(\e^2) \quad \mbox{for}\ E,T\;,
\end{equation}
in agreement with Eqs.~(\ref{avalues}), (\ref{thvalue}) and the
scaling forms~(\ref{scalformCfin}) and~(\ref{scalformRfin}), both for
$E$ and $T$.

\end{document}